%% file: main.tex
\documentclass[fleqn,usenatbib]{mnras}

\usepackage{newtxtext,newtxmath}

\usepackage[T1]{fontenc}

\DeclareRobustCommand{\VAN}[3]{#2}
\let\VANthebibliography\thebibliography
\def\thebibliography{\DeclareRobustCommand{\VAN}[3]{##3}\VANthebibliography}

\usepackage{graphicx}	
\usepackage{amsmath}	
\usepackage{subcaption}

\usepackage{csvsimple} 
\usepackage{longtable} 
\usepackage{booktabs}
\usepackage{multicol}
\usepackage{lscape}
\usepackage{caption}

\newcommand{\approximately}{\raisebox{0.5ex}{\texttildelow}}

\title[Exoplanet Transit Candidate Identification]{Exoplanet Transit Candidate Identification in TESS Full-Frame Images via a Transformer-Based Algorithm}

\author[Helem Salinas et al.]{Helem Salinas$^{1,2}$\thanks{E-mail: yhsalinas@uc.cl}, 
Rafael Brahm$^{4,5,6}$, 
Greg Olmschenk$^{3,7}$\thanks{ORCID: 0000-0001-8472-2219},
Richard K. Barry$^{3}$\thanks{ORCID: 0000-0003-4916-0892},
Karim Pichara$^{1,5}$,
\newauthor Stela Ishitani Silva$^{3,8}$\thanks{ORCID: 0000-0003-2267-1246},
Vladimir Araujo$^{9}$ \\
\\
$^{1}$Departamento de Ciencias de la Computación, Facultad de Ingeniería, Pontificia Universidad Católica de Chile, Santiago 7820436, Chile\\
$^{2}$Department of Physics, The Catholic University of America, Washington, DC 20064, USA\\
$^{3}$NASA Goddard Space Flight Center, Greenbelt, MD 20771, USA\\
$^{4}$Facultad de Ingeniería y Ciencias, Universidad Adolfo Ibáñez, Av. Diagonal Las Torres 2640, Peñalolén, Santiago, Chile\\
$^{5}$Millennium Institute for Astrophysics, Av. Vicuna Mackenna 4860, 782-0436 Macul, Santiago, Chile\\
$^{6}$Data Observatory Foundation, Chile\\
$^{7}$Department of Astronomy, University of Maryland, College Park, MD 20742, USA\\
$^{8}$Oak Ridge Associated Universities, Oak Ridge, TN 37830, USA\\
$^{9}$Department of Computer Science, KU Leuven, Celestijnenlaan 200A, B-3001 Leuven, Belgium
}

\date{Received 2025 January 16; Revised 2025 February 10, Accepted 2025 February 16}

\begin{document}
\label{firstpage}
\pagerange{\pageref{firstpage}--\pageref{lastpage}}
\maketitle

\input{abstract}


\input{introduction}

\input{background_theory}

\input{data}

\input{method}


\input{training_process}

\input{catalogue}

\input{conclusions}

\section*{Acknowledgements}
The authors would like to acknowledge the support from the National Agency for Research and Development (ANID), through the FONDECYT Regular project number 1180054. HS acknowledges support from ANID, through Scholarship Program/Doctorado Nacional/2020-21212185. R.B. acknowledges support from FONDECYT Project 1241963 and from ANID -- Millennium  Science  Initiative -- ICN12\_009. GO's work is supported by NASA under award number 80GSFC21M0002. SIS’s research was supported by an appointment to the NASA Postdoctoral Program at the NASA Goddard Space Flight Center, administered by Oak Ridge Associated Universities under contract with NASA. This paper utilizes public information of the TESS mission. Funding for the TESS mission is provided by the NASA Explorer Program. Finally, thanks to the availability of data from TESS, we were able to examine and display our analysis of transit signals.

This research has made use of the Exoplanet Follow-up Observation Program (ExoFOP; DOI: 10.26134/ExoFOP5) website, which is operated by the California Institute of Technology, under contract with the National Aeronautics and Space Administration under the Exoplanet Exploration Program.

\section*{DATA AVAILABILITY}

All the data referenced here and that was used is public information about the TESS mission from Mikulski Archive for Space Telescopes (MAST), as well as data from the European Space Agency (ESA) mission Gaia. 

\bibliographystyle{mnras}
\interlinepenalty=10000

\bibliography{main}

\appendix
\section{Multi-Head Self-Attention}
\label{sec:appendix_att_mechanism}
\input{appendix}





\bsp	
\label{lastpage}
\end{document}

%% file: abstract.tex
\begin{abstract}

The Transiting Exoplanet Survey Satellite (TESS) is surveying a large fraction of the sky, generating a vast database of photometric time series data that requires thorough analysis to identify exoplanetary transit signals. Automated learning approaches have been successfully applied to identify transit signals. However, most existing methods focus on the classification and validation of candidates, while few efforts have explored new techniques for the search of candidates. To search for new exoplanet transit candidates, we propose an approach to identify exoplanet transit signals without the need for phase folding or assuming periodicity in the transit signals, such as those observed in multi-transit light curves. To achieve this, we implement a new neural network inspired by Transformers to directly process Full Frame Image (FFI) light curves to detect exoplanet transits. Transformers, originally developed for natural language processing, have recently demonstrated significant success in capturing long-range dependencies compared to previous approaches focused on sequential data. This ability allows us to employ multi-head self-attention to identify exoplanet transit signals directly from the complete light curves, combined with background and centroid time series, without requiring prior transit parameters. The network is trained to learn characteristics of the transit signal, like the dip shape, which helps distinguish planetary transits from other variability sources. Our model successfully identified 214 new planetary system candidates, including 122 multi-transit light curves, 88 single-transit and 4 multi-planet systems from TESS sectors 1-26 with a radius > 0.27 $R_{\mathrm{Jupiter}}$, demonstrating its ability to detect transits regardless of their periodicity.

\end{abstract}

%% file: introduction.tex
\section{Introduction}

The amount of astronomical data collected from both space and ground-based telescopes has increased rapidly. Given the vast amount of data, the task of analysis and examination becomes quite costly and is also prone to human error. Process automation plays an important role in helping to improve the efficiency and accuracy of detection and classification. \par

During the last decade, more than a million stars have been observed in the search for transiting exoplanets. Among the most notable contributors is NASA’s Kepler space telescope, launched in 2009 \citep{borucki2010kepler}. Kepler has identified nearly 4,000 potential planet candidates, of which approximately 2,700 have been confirmed as exoplanets to date. Kepler's successor, NASA's Transiting Exoplanet Survey Satellite \citep[TESS;][]{ricker2014transiting} was launched in April 2018 and is currently monitoring most of the sky searching for transiting exoplanet candidates. To date, TESS has discovered 561 confirmed exoplanets and identified thousands of additional planet candidates, significantly expanding our understanding of planetary systems beyond our solar system. TESS monitors millions of stars across approximately 90\% of the sky, emphasizing the detection of exoplanets around nearby and bright stars. TESS observes each sector for \approximately{}27 days before reorienting to observe the next sector. The data products produced by the TESS mission include two types of data: small summed image subarrays (``postage stamps'') centred on pre-selected targets, also known as target pixel files (TPFs), and summed full-frame images (FFIs), which capture collections of pixels observed simultaneously \citep{guerrero2021tess}. From these data products, the primary goal of TESS is to discover hundreds of transiting planets smaller than Neptune, including dozens comparable in size to Earth \citep{Ricker:2015}. \par

The current standardized identification process of transiting exoplanet candidates from photometric time series involves two main steps. First, long-term flux variations are removed, and periodic transit-like signals are identified by phase-folding the light curve to different orbital periods, times of transit and transit duration \citep[e.g.][]{tenenbaum:2012}. This process yields Threshold Crossing Events (TCEs), which include both potential exoplanets and false positives like eclipsing binaries and instrumental artifacts. The second step involves vetting TCEs by rejecting false positives based on the properties of the phase-folded transiting light curves, stellar properties, and follow-up observations. Typically, experts manually examine possible exoplanet transit signals, a labour-intensive process prone to human error and increasingly demanding due to large data volumes. To address this, several efforts have been made to automate the classification of light curves. \par

Meanwhile, with the advent of Deep Learning (DL) \cite {lecun2015deep}, neural network (NN) architectures have revolutionized numerous fields of scientific research, leading to substantial advancements in image processing, classification, facial recognition, voice recognition, and Natural Language Processing \cite[NLP;][]{krizhevsky2017imagenet, ciregan2012multi, he2016deep, voulodimos2018deep, vaswani2017attention}. DL techniques, particularly convolutional neural networks \citep[CNN;][]{krizhevsky2012imagenet}, have been notably applied to classification tasks, where classification in machine learning refers to the task of predicting the class to which the input data belongs. In particular, it was used to classify transit signals for the validation of exoplanet candidates and has also been applied in the transit search process, though to a smaller extent. In the field of exoplanet transit signal classification, 1D CNNs have been commonly used \citep[e.g.][]{osborn2020rapid, rao2021nigraha, tey2023identifying}. Specifically, the approaches focus on validating transit signals obtained through phase-folding light curves, which depend on prior transit parameters such as period, duration and depth derived from initial detection processes. These methods utilize both local and global views of folded light curves for distinguishing exoplanet transits from false positives, but they do not perform the initial transit search or parameter extraction themselves. Regarding the search for new exoplanet candidates, \cite{olmschenk2021identifying} developed a pipeline for detecting and vetting periodic transiting exoplanets. Their approach utilizes the TESS FFIs light curves and employs a 1D-CNN for candidate identification. Another approach, as demonstrated by \cite{cui2021identify}, employs a 2D object detection algorithm based on YOLOv3 network \citep{redmon2018yolov3incrementalimprovement}, which was trained on Kepler data to detect exoplanet signals. \par

As mentioned above, proposed DL pipelines have primarily focused on classifying transiting exoplanet signals, where transit detections are confirmed by aligning several repeated transits with the orbital period, resulting in enhanced signal detection. In contrast, few efforts have explored new techniques for discovering candidates without requiring prior transit parameters. Consequently, while current standardized pipelines for detecting transiting exoplanets have reached a high level of efficiency, they have at least two important limitations, namely: i) they require de-trending processes to remove long-term flux variations of stellar origin, which could vary the trend of the sequence, introduce artifacts into the light curve, or remove exoplanet transits; and ii) they rely on the periodicity of the transit signal in the light curve. These pipelines could be therefore missing exoplanets that present strong transit timing variations (TTV) and/or those that present just one transit in the light curve (single transiters). \par

Regarding single transiters, many signals in observed light curves arise from stochastic processes—whether instrumental or astrophysical—which can often mimic single transit events \citep{foreman2016population}. To accurately determine the orbital period of this single transit, detecting additional transits in multiple sectors is necessary, given that interruptions in the light curves typically require at least three transit events to reliably constrain the orbital period. This is because the presence of a single transit event does not provide sufficient information to confirm the periodic nature of the signal. This leads to considerable challenges that hinder a search for single transit events. For instance, the transit probability for long-period planets could be lower, particularly for those with orbital periods that exceed the duration of continuous observations \citep{hawthorn2024tess}. This is also due to the fact that transit probability $P_{\mathrm{tr}}$ is proportional to ${r_\ast}/a$ and for long-period planets, the semi-major axis $a$ is larger, resulting in a lower $P_{\mathrm{tr}}$. Despite the challenges associated with detecting single transiters, a small number of named ``monotransiters'' have been confirmed, where a single transit-like feature appears in the light curve but does not repeat \citep{gill2020ngts1, gill2020ngts2, lendl2020toi}.  \par


Due to observational biases, the vast majority of well-characterized transiting giant planets have orbital periods shorter than 10 days [e.g.]\citep{hartman:2019, magliano2023tess}. The existence of these planets, known as hot Jupiters, has challenged theories regarding the formation and evolution of giant planets in general \citep{pollack:96}. To solve these open questions, it is crucial to discover and characterize giant planets with longer periods (warm Jupiters) that are not extremely irradiated by their host stars, and their physical and orbital properties can be compared to outcomes of different formation and migration processes \citep{dawson:2018}. While TESS has significantly increased the sample of well-characterized transiting warm Jupiters \citep[e.g.][]{dawson:2021,grieves:2022,brahm:2023,battley:2024}, an important fraction of them might still be undiscovered because they are presented as single transiters in TESS data \citep[e.g.][]{gill:2020}. \par

In this work, we propose an approach to identify exoplanet transit signals within a light curve without requiring prior transit parameters. This enables us to discover new exoplanet candidates directly from the complete light curve, avoiding the initial dependence on traditional techniques that phase-fold the light curve to identify periodic transit-like signals. We propose an architecture based on a sequential model inspired by the Transformer \citep{vaswani2017attention}. This type of architecture has demonstrated its effectiveness in handling complex scenarios, outperforming other model types such as recurrent neural networks \citep[RNN;][]{cho2014learning} and CNNs, in particular for sequential data \citep{hawkins2004problem, lakew2018comparison, karita2019comparative}. Our architecture learns the characteristics of a planetary transit signal, regardless of whether the signal is periodic throughout the light curve. While instrumental systematics are typically removed during preprocessing, to compute light curves, stellar-origin variations remain in the data. Our NN is therefore exposed to both long-term and short-term stellar variability, including activity-related fluctuations on timescales of days. Specifically, the architecture is designed to identify and differentiate between transit signals and the variability of stars. This type of architecture has been explored and applied by \cite{10.1093/mnras/stad1173} for classifying exoplanet transit signals and distinguishing them from false positives, and our NN extends this approach to the search for new exoplanet candidates. In particular, our architecture is designed to make predictions directly from the light curve. The NN captures temporal patterns and long-range dependencies within light curves without depending on prior transit parameters, such as transit depth, transit period and transit duration. This independence from periodicity allows our NN to detect both multi-transit and single-transit signals, making it particularly effective for identifying single transiters that have often been missed by previous approaches focused primarily on light curves with multiple transits of the same planet. \par


This article is organized as follows: Section~\ref{sec:background} introduces the key concepts that inspired this work, where we briefly describe the Transformer encoder related to the original architecture \citep{vaswani2017attention} in order to understand our proposal for exoplanet transit signal identification. Section~\ref{sec:dataset_tess} describes the photometric data used in our work. Section~\ref{sec:ourmethod} explains the proposed methodology in detail, including the NN training process. The model analysis and results are described in Section ~\ref{sec:model_analysis}. In Section ~\ref{sec:search_candidates}, we describe the exoplanet candidates found. Finally, in Section~\ref{sec:conclusions} we present overall conclusions and future work.

%% file: background_theory.tex
\section{Background Theory of Transformer Encoder}
\label{sec:background}


Recent developments in DL have introduced new approaches to processing sequential data. Among these, the Transformer model proposed by \cite{vaswani2017attention} has recently demonstrated significant success in capturing long-range dependencies compared to previous approaches focused on sequential data \citep{hawkins2004problem, lakew2018comparison, karita2019comparative}.  Initially developed for Natural Language Processing, it has since been adapted to various domains, including time-series analysis. In NLP, data typically consists of discrete tokens, such as words or characters, arranged in a sequence to form sentences or documents. In astronomy, by contrast, the data often comes in the form of continuous measurements, such as light curves that represent stellar brightness over time. (In the present time-series context, for example, the term range refers to the length of the time baseline between separate elements within the data.) Both are sequential data, where the order of elements within the sequence is essential to uncovering patterns and relationships. \par

The original Transformer model consists of an encoder and a decoder \citep{vaswani2017attention}. In this work, we focus on the encoder. The encoder processes the input sequence to generate a representation, which captures the essential patterns and features of the data. \par

The core of the Transformer encoder is a mechanism called self-attention, which builds upon the original attention mechanism proposed by \cite{bahdanau2014neural}. Self-attention enables the NN to ``attend'' to all parts of the input sequence simultaneously. This mechanism allows the NN to evaluate the importance of each element in the sequence in relation to every other element, weighing how much each data point should contribute to the final representation. It enables the NN to capture both local and long-range dependencies. Unlike traditional methods that focus on short-range dependencies, self-attention allows the Transformer to understand the full temporal structure of the data \citep{lakew2018comparison, karita2019comparative}. \par

\begin{figure*}
\begin{center}
\vspace{5mm}
\includegraphics[width=0.87\textwidth]{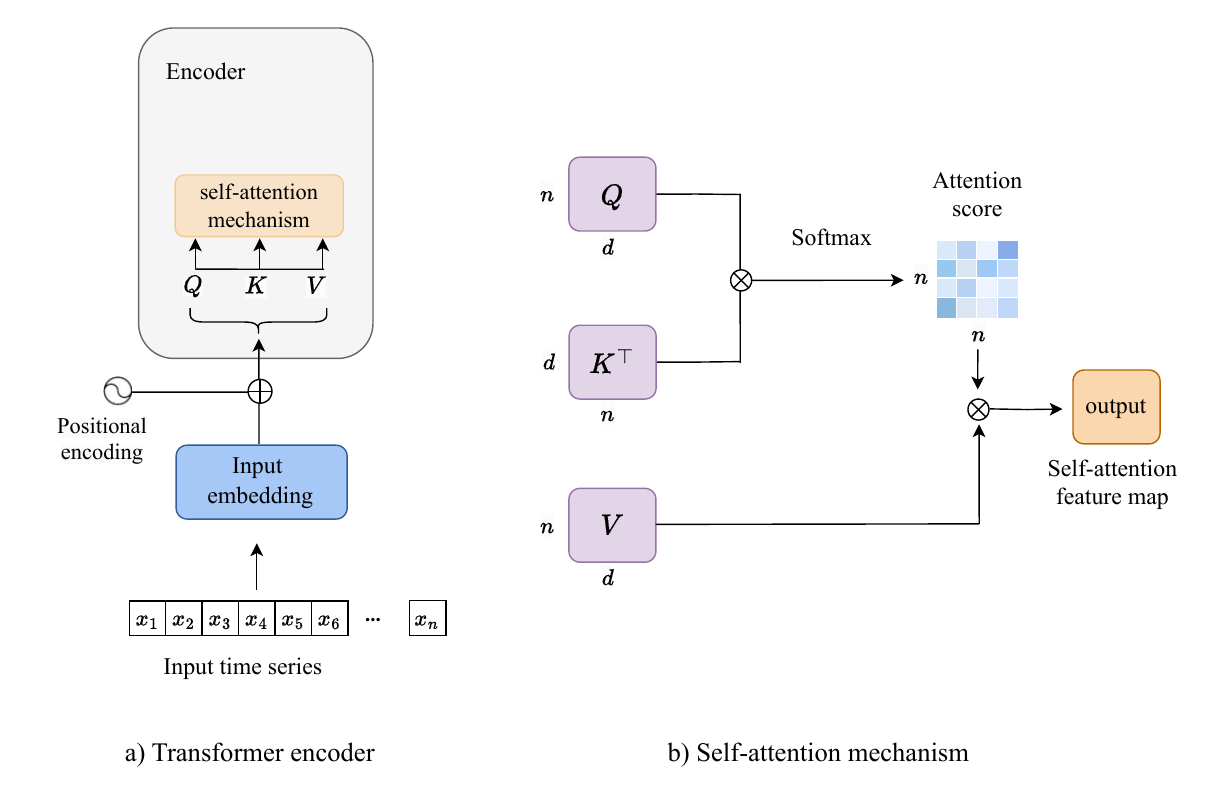}
\caption{\label{fig:self-attention}
(a) the transformer encoder, which processes time series inputs using positional encodings combined with input embeddings, and computes feature representations through a self-attention mechanism. (b) explains the self-attention mechanism, where the $Q$, $K$ matrices are used to calculate attention scores. These scores are then applied to the $V$ to produce the self-attention feature map.}
\vspace{-1mm}
\end{center}
\end{figure*}

Figure \ref{fig:self-attention} shows the core of the Transformer encoder, and provides a detailed view of the self-attention mechanism. The input sequence is represented as a series of data points, denoted as $(x_1, x_2, \ldots, x_n)$, where $n$ is the length of the sequence. This input is first encoded into a continuous vector space through the embedding layer, and each $x_i$ is represented as a vector $x_{\mathrm{emb}_i} \in \mathbb{R}^d$, where $d$ is the embedding dimension. Vector embedding is a way to represent data symbols (such as words, observations, or images) in a continuous, transformed space. Embeddings can be numeric representations or generated through NNs. As shown in Figure \ref{fig:self-attention}(a), this representation is then combined with a positional encoding, which incorporates information about the order of the data points, capturing their relationships in the temporal dimension \citep{vaswani2017attention}, such as patterns and relationships that depend on the temporal order of the data, such as periodic trends or sequential dependencies. The positional encoding is essential because the Transformer architecture, unlike RNN or CNN, does not have an inherent way to understand the order of elements in a sequence \citep{gehring2017convolutional}. This combined representation is then passed to the encoder. Within the encoder, the self-attention mechanism captures both local and global dependencies, enabling the extraction of complex features and patterns.  \par

The self-attention mechanism used within the encoder to process the input sequence, in this case a light curve. The self-attention process involves three components: queries ${Q}$, keys ${K}$ and values ${V}$. As shown in Figure \ref{fig:self-attention}(b), each element in the sequence generates a query, which asks how much attention it should pay to the other elements. The query is compared to keys, which represent the context, and an attention score is calculated. The score is calculated using the scaled dot-product of ${Q}$ and the transpose of ${K}$. These values are then passed through a softmax function to convert the values into a probability distribution that sums to 1, while sharpening the distribution. This score determines how much focus each element should give to others. \par

Finally, ${V}$, weighted by the attention scores, are then combined to produce the final output of the self-attention feature map. This feature map encapsulates the relationships across the sequence and contributes to the encoded representation $z$. In the case of multihead self-attention (MSA), multiple attention mechanisms are applied in parallel, capturing different aspects of the relationships across the sequence (see the Appendix \ref{sec:appendix_att_mechanism} for more details). The decoder then uses this encoded representation $z$ to produce outputs such as probability estimates for the presence of planetary transit signals into a light curve. This approach enables the model to detect exoplanet transit signals directly from the complete light curve without prior transit parameters even if only one transit is exhibited during the entire observation sequence.  \par

Transformers have shown significant potential for encoding time series data in astronomy \citep[e.g.][]{allam2021paying, zerveas2021transformer, morvan2022don}. In the field of exoplanet science, \cite{10.1093/mnras/stad1173} implemented a Transformer model specifically tailored to distinguish exoplanet transit signals from false positives, highlighting the capability of the architecture to identify and interpret the characteristics of exoplanet signals. In the context of exoplanet transit signal detection, the ability of the Transformer to model long-range dependencies is highly beneficial, as it can identify signal patterns associated with stellar variability that might otherwise be mistaken for false positives. This allows the model to accurately distinguish true planetary transits within the entire light curve, even amid the variability of the stars.  \par

%% file: data.tex
\section{TESS Data}
\label{sec:dataset_tess}

For the present work, we have used light curve data obtained from the Mikulski Archive for Space Telescopes (MAST). The light curves are extracted from the TESS full-frame images (FFIs), which were captured at a cadence of 30 minutes during its primary mission. The time stamps for these measurements are given in Barycentric TESS Julian Date (BTJD) format, which is specific to the TESS mission. BTJD is defined as $\mathrm{BTJD = BJD - 2457000.0}$ days, where BJD refers to the Barycentric Julian Date. The use of BTJD ensures precise timing by taking into account the motion of Earth and other relativistic effects. To obtain the light curves, we use the outputs of the TESS Science Processing Operations Center (SPOC) pipeline \citep{caldwell2020tess} (see Section~\ref{sec:tess_spoc_ffi}). SPOC pipeline processes the data of targets selected from the FFIs to create target pixel and light curve files for up to 150,000 targets per sector. By selecting a specific set of target stars from these FFIs, SPOC generates the same outputs as for the pre-selected 2-min cadence targets, including calibrated target pixel files and simple aperture photometry (SAP) flux.  \par


\subsection{TESS-SPOC FFI light curves}
\label{sec:tess_spoc_ffi}

For the primary mission of TESS, the FFIs are captured at a 30-minute cadence and are crucial for large-scale surveys, enabling the simultaneous monitoring of thousands of stars. To optimize the utility of FFI data, SPOC employs a selection process that prioritizes targets based mainly on their crowding metric \citep{bryson2020kepler} and brightness. Specifically, targets with a crowding metric of $\geq$ 0.5 are selected, ensuring that at least 50\% of the flux within the photometric aperture originates from the target star itself. Additionally, SPOC focuses on TESS Input Catalog (TIC) objects with TESS magnitudes of T $\leq$ 16, balancing the need for both bright and isolated targets to maximize the scientific return from the FFI data \citep{caldwell2020tess}. Building on this, our architecture is trained with the light curves processed by SPOC from TESS sectors 1-26. Each sector was observed continuously for about 27 days, with FFIs captured at a 30-minute cadence. \par

The light curves computed by SPOC from the FFIs include the stellar flux (PDCSAP\_FLUX), which has been corrected for instrumental systematics and shows fewer systematic trends. In addition to the flux, SPOC also provides detailed information on the centroid position of the target star and the background flux. The centroids are a time series that represents the pixel position of the centre of the light, which varies throughout the observation. This information allows us to determine the location of the transit within the pixels. When the centroid shifts during a transit event, we can identify whether the observed flux variations are coming from the target star or from nearby sources. The background time series represents the estimated flux contribution from nearby sources, such as nearby stars or unresolved objects, to the target aperture. It is derived from pixels located outside the photometric aperture but within the same CCD frame. This estimation aims to minimize the influence of nearby stars, though it is not always possible to completely exclude their effect, but their contribution is minimized. Both time series allow the distinction between true transits and other astrophysical phenomena that might affect the measured light curve caused by background contamination or blended stellar sources. We use these data sets; PDCSAP\_FLUX, centroids, and background time series as inputs to our architecture, enhancing the accuracy and reliability of exoplanet transit detection. \par


%% file: method.tex
\section{Methodology}
\label{sec:ourmethod}

In this section, we describe our methodology, which is designed to determine the confidence level of potential transiting exoplanets within a given FFI light curve. Our work is mainly motivated by the need to detect exoplanet candidates observed by the TESS mission. To determine the confidence level, our proposed architecture considers the flux, centroid, and background time series as input of our NN. Our methodology consists of three main steps, which we describe in this section. First, in Section~\ref{sec:preprocessing}, we detail the data preprocessing steps used to construct our input representation, such as the time series and the data augmentation techniques employed. Section~\ref{sec:ground_truth} discusses the ground truth data utilized in our analysis. Finally, Section~\ref{sec:nn_architecture} presents the design of our proposed NN architecture, with a description of the experimental setup and training details.

\subsection{Data preprocessing}
\label{sec:preprocessing}
The preprocessing of the light curves provided by SPOC prepares them for model input through necessary transformations. In this process, we ensure that the time series are in a consistent format, allowing the NN to process and detect significant patterns within the light curves. \par

\subsubsection{Flux, centroid, and background time series}

Each light curve is processed to obtain a uniform length of 1000 data points, reflecting the approximate median length of single-sector light curves derived from TESS FFI data. Light curves longer than 1000 data points are truncated to this length. And, for light curves with fewer data points, we extend them by repeating the initial segment until they reach 1000 points. This standardization allows the same sequence sizes for the input for our model. Following length adjustment, we normalize the flux time series values. Specifically, we subtract the mean and divide by the standard deviation, and these values are linearly transformed to a range of -1 to 1. \par

We also incorporate two more time series. First, we included the centroid information into our input. For this, we use the CCD row and column positions of the target centroid provided in the SPOC dataset, with $x_{\mathrm{row}}$ representing the row position and $y_{\mathrm{col}}$ representing the position of the column. To characterize the centroid displacement, we compute its absolute magnitude as $r_{\mathrm{centroid}} = \sqrt{x^2_{\mathrm{row}} + y^2_{\mathrm{col}}}$. Then, the centroid time series is then normalized to the range -1 to 1, following the same procedure as the flux. Finally, we include background time series from the light curves, which is also normalized in the same way as the flux and centroid time series. This ensures that the background data, along with flux and centroid information, are on a comparable scale to maintain consistency across all input features.  \par


\subsubsection{Augmentation of training data}
\label{sec:augmentation}
Through various augmentation methods, the NN is allowed to learn more generalized features and prevent network overfitting. We applied four techniques, exclusively to the training dataset. First, we add white noise to the flux values with a standard deviation randomly chosen from a uniform distribution between 0 and the mean of the flux. Second, we use random roll augmentation, which cyclically shifts the time series data for flux, centroid, and background by a random number of positions. This shift generates variations in the data without altering the core characteristics of the signal, thus preventing the model from overfitting to specific patterns. Third, we apply random split and swap augmentation, where the time series data is split at a random index, and the resulting segments are swapped, maintaining temporal relationships while generating new variations. As fourth and last, mirror augmentation is used, which involves flipping the time series data along the time dimension. By reversing the order of data points, this technique creates a mirrored version of the original series, introducing further variations while preserving the overall patterns and relationships within the data. \par


As mentioned before, these augmentation methods are specifically applied to the training data and not to the validation or test datasets, ensuring that the model performance evaluation remains unbiased and reflective of its generalization capabilities. Specifically, the augmentation technique applied to each sample is randomly selected in every iteration, meaning that the model always sees different variations of the data. The stochastic nature of this augmentation strategy allows the model to be exposed to varying patterns and conditions, making it less likely that the model will memorize specific patterns present in the training data. This approach not only broadens the range of scenarios the model is trained on but also significantly enhances its robustness and ability to generalize to new, unseen data \citep{shorten2019survey}. This helps the model to perform optimally during the inference phase. \par


\subsection{Ground truth data set}
\label{sec:ground_truth}

For our ground truth dataset, we collected samples for both positive and negative cases. We categorized the collected labels into three primary sets of light curves: known transiting planets, eclipsing binaries (EBs), and non-transit signals. The first set, representing positive cases, was drawn from the TESS Exoplanet Follow-up Program (ExoFOP-TESS) catalog\footnote{https://exofop.ipac.caltech.edu/tess/}, which provided labels for light curves corresponding to confirmed and known planets. This dataset was utilized for training, validation, and testing of our NN. It is important to mention that we excluded any unconfirmed planet candidates listed by ExoFOP-TESS, as many of these are later identified as false positives during follow-up observations. \par

The second set consisted of EBs collected from three main catalogues. First, we used the catalogue from \cite{prvsa2022tess}, which provides a collection of EBs observed by TESS. Second, we incorporated false positives classified as EBs from the ExoFOP-TESS catalogue. These targets were initially identified as planetary candidates but were later reclassified as EBs following detailed follow-up observations. Finally, we added a catalogue based on ML predictions from \citet{yu2019identifying}(referred as ML:Y19 in Table \ref{tab:ds_gt}), which includes labelled EBs and is publicly available\footnote{Available at \url{https://github.com/yuliang419/Astronet-Triage}}. EB contamination are among the most common false positives due to their periodic patterns, which are similar to true exoplanet transits. These strong negative cases were crucial for training the model to distinguish between true transits and false positives. \par

The third set consists of non-transit signals derived from a catalogue based on machine learning predictions from \citet{yu2019identifying}. This dataset included labels for instrumental systematics (IS), stellar variability (V), and a Junk (J) class, which represents a mix of IS and V signals. With these non-transit light curves, we expanded the variety of negative cases. In addition, we included light curves labelled as false positives (FP) by the ExoFOP-TESS catalogue, ensuring that the selected false positives were not associated with EBs. \par

Since each target may have been observed in multiple sectors, and exoplanet transits appear in specific sectors based on their periodicity, we filtered the light curves to include only those that captured the exoplanet transits. The same approach was applied to EBs, where we selected the light curves containing EB transits. This approach has the advantage of capturing the unique noise characteristics from each sector, as each sector introduces its own variations, which help train the model to recognize signals under different observational conditions. Table~\ref{tab:ds_gt} provides a detailed summary of the light curves collected for each target. \par


Additionally, we trained our network using light curves artificially injected, following a process inspired by \cite{olmschenk2021identifying}. We injected signals from one light curve into another, selected from the three primary sets of light curves described above. This involved injecting light curves containing transit signals from confirmed exoplanets into non-transit light curves. This approach allows the network to learn from a variety of signals, incorporating real stellar variability and noise conditions. To achieve this, we randomly selected a light curve from the base set and normalized it by applying a median scaling to produce a relative magnification signal. Next, we randomly sampled a non-transit light curve and scaled its values by the previously generated relative magnification signal. We repeated this process, injecting light curves from EBs into non-transit light curves. Thus, light curves injected with known or confirmed exoplanets signals were labelled as positive ground truth, while those injected with EBs signals were labeled as negative ground truth. Non-transit light curves remained as negative ground truth, ensuring they continued to represent non-planetary signals. \par

Our training setup involves showing the network three types of light curves in each iteration. Specifically, we ensured that each batch contained an equal number of light curves from planets, EBs, and non-transit signals. This balanced distribution forces the network to learn and differentiate between exoplanet transit signals, EB signals, and other false positives, improving its ability to distinguish each type of signal. \par

The ground truth dataset was divided into training, validation, and testing subsets. Specifically, 80\% of the dataset was designated for training, 10\% for validation, and 10\% for testing. The testing subset provides a final assessment to measure the model's generalization capabilities across different light curves and transit signals. \par


\begin{table}
	\centering
	\vspace{5mm}
	\caption{Labels and datasets used to train and evaluate our model.}
	\label{tab:ds_gt}
	\vspace{4mm}
	\begin{tabular}{lcccc} 
		\hline
		Label & Target & Catalogue  & Total  \\
		 &  & & sectors 1-26  \\
		\hline
		Positive & confirmed planet & ExoFOP &   1230\\
        \hline
		Negative & EBs & TESS-EBs &  9333 \\
                 & & ExoFOP &  1049 \\
                 & & ML:Y19 &  6550 \\
        \vspace{0.1mm}\\
        \cline{2-4} 
        \vspace{0.1mm}\\
        & Non-transit & ExoFOP &   934  \\
        &  & ML:Y19 &   26200 \\

		\hline
		\vspace{1mm}
	\end{tabular}
\end{table}

\subsection{Neural network architecture}
\label{sec:nn_architecture}

\begin{figure*}
\begin{center}
\vspace{5mm}
\includegraphics[width=0.6\textwidth]{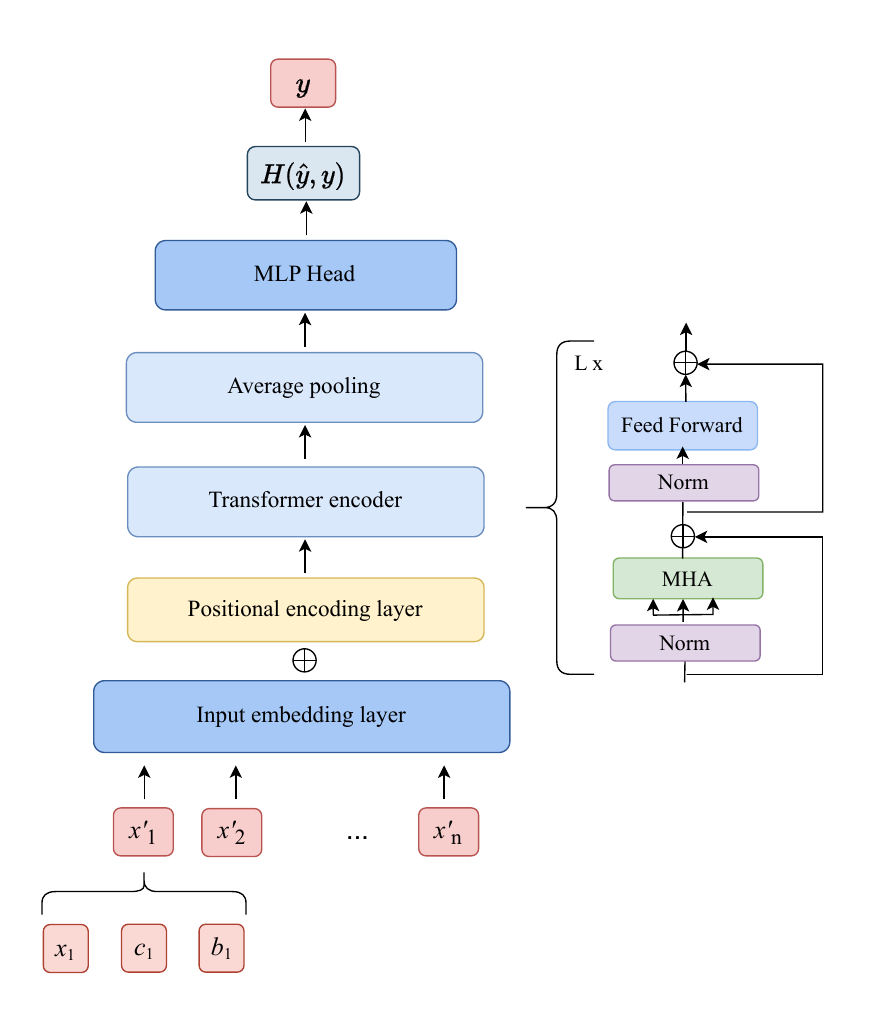}
\vspace{2mm} 
\caption{\label{fig:architecture}
Schematic of the proposed architecture. The input includes flux $x_i$, centroid $c_i$, and background $b_i$ time series,  which are concatenated into input embeddings $x'_i$ which are processed using convolutional embeddings. The tokens, along with positional encodings, are passed through a MSA mechanism within a transformer encoder. The features embedding produced by the transformer encoder are then passed through to average pooling, followed by a feed-forward MLP head to predict the class. The predicted output $\hat{y}$ is evaluated using a loss function $H(\hat{y}, y)$ to classify the input into one of two classes(0 or 1).}
\vspace{-1mm}
\end{center}
\end{figure*}

Our proposed architecture is illustrated in Figure~\ref{fig:architecture}, which integrates convolutional embeddings with a Transformer encoder to process light curve data. The process begins with convolutional embedding, which operates on raw inputs such as flux, centroid, and background, extracting local features through localized patterns in the data. These embeddings are then combined with positional encodings, which inject information about the positions of data points in the sequence. Since transformer-based NN do not inherently process sequential information, positional encodings help the model understand the order of the data, enabling it to capture temporal dependencies and the relative positioning of data points \citep{vaswani2017attention}. The combined sequence of embeddings is then passed into the Transformer encoder. Then, the transformer encoder outputs are passed through average pooling to condense the information, followed by a multi-layer perceptron (MLP) head, which maps the features into a classification space. This final stage produces the probability of the input light curve representing a light curve with exoplanet transit (class 1) or a light curve without exoplanet transit (class 0). The combination of convolutional embeddings for local feature extraction and Transformer encoders for global contextualization enables the model to robustly identify exoplanet signals, even when faced with complex stellar variability. \par

\begin{figure}
\begin{center}
\vspace{5mm}
\includegraphics[width=0.49\textwidth]{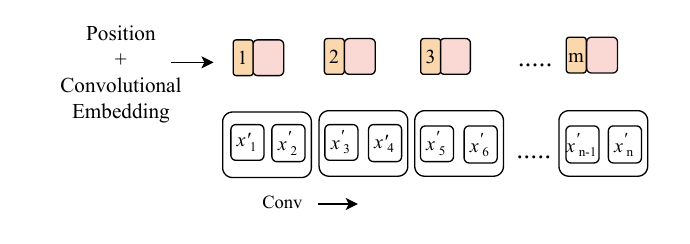}
\vspace{-2mm} 
\caption{\label{fig:cnn_embedding}
CNN embedding, where the kernels slide across the input time series, transforming each local window into an embedding vector. }
\vspace{-1mm}
\end{center}
\end{figure}

\subsubsection{Tokens and convolutional embedding}
\label{sec:inputembedding}
The input time series data, including flux, centroid, and background, are first segmented into tokens, representing key features within the series.
The first step for each piece of information is to generate an input sequence, for this, we generate a sequence of observations in terms of the flux, centroid and background time series, represented as $\{(x_1, c_1, b_1), (x_2, c_2, b_2), (x_3, c_3, b_3), ..., (x_n, c_n, b_n)\}$, where $x_i, c_i, b_i$ denote the flux, centroid, and background values, respectively, at each time step $i$. Here, $n$ is the length of time steps in the sequence. Each concatenated vector is defined as: 

\vspace{-2.5mm}
\begin{equation}
x'_i = (x_i, c_i, b_i)
\end{equation}

where $x' \in \mathbb{R}^m$, and $m$ is the number of variables of our time series. After concatenation, the sequence of these concatenated vectors $ X = \{x'_1,x'_2, ..., x'_n\}$ is the input for the convolutional operation.

As second step, we build the model's vector embeddings. To obtain this embedding, we apply two layers of 1D convolution to our input sequence $X$. Each resulting embedding vector, or token, captures essential information about the input signal. Figure \ref{fig:cnn_embedding} shows this embedding process, where each input element is processed by CNN kernels that slide across the time series, transforming each local window into a token. These tokens extract and capture local patterns and short-term dependencies within the convolution window to identify patterns of exoplanet transit signals. In this context, the CNN focuses on the time variations in brightness that occur over short intervals, allowing the model to detect the distinct shape of the exoplanet transit signal as it passes in front of its host star. \par

We denote the convolution filter coefficients as $\mathbf{w} \in \mathbb{R}^{n \times k}$, and the output of the convolution operation at position $i$ is the embedding token $v'_i$ which is computed as:

\begin{equation}
v'_i = \mathbf{w}^{\left ( k \right )}_i \cdot x'_{i} = \mathbf{w}^{\left ( k \right )}_i \cdot (x_i, c_i, b_i)
\label{eq:conv_embb}
\end{equation}

In the first CNN layer, the kernel size $k=1$ slides, transforming an $m$-dimensional space (representing the concatenated values of flux, centroid, and background) into a $d$-dimensional space at each time step $t$. The output of the first CNN layer is the input of a second convolution. The number of filters in the second convolution layer is set to match the model dimension $d$. The final embedding after the second convolution layer is expressed as: \par

\begin{equation}
\mathbf{X}_{\text{emb}} = \text{CNN}_2\left( \text{CNN}_1\left( X \right) \right)
\end{equation}

where, $\mathbf{X}_{\text{emb}} \in \mathbb{R}^{T \times d}$ is the embedded representation, with $T$ being the number of tokens generated after the convolution process, $X$ represents the input features (flux, centroid, background) and $d$ is the dimension of the output embedding space.

After the layer where we create the convolutional embedding, we combine it with the positional encoding method proposed by \cite{vaswani2017attention}. Positional encodings provide information about the order of the time steps; it allows the model to identify temporal dependencies and relationships in the data after the CNN layers. The positional encoding is added to the embeddings as: \par

\begin{equation}
\mathbf{Z}_{0} = \mathbf{X}_{\text{emb}} + \mathbf{P}
\label{eq:z_embedding}
\end{equation}


where $\mathbf{P} \in \mathbb{R}^{T \times d}$ represents the positional encoding for each token to help the model interprets the sequential pattern, which is defined as:

\begin{equation}
\mathbf{P}_{\mathnormal{(p, 2i)}}=\mathrm{sin}\left ( \frac{p}{10000^{2i/d}} \right ),  \quad  \mathbf{P}_{\mathnormal{(p, 2i+1)}}=\mathrm{sin}\left ( \frac{p}{10000^{2i/d}} \right )
\label{equation_pe}
\end{equation}

where $p$ represents the position of an element in the input sequence and $i$ that is used for mapping to indices of the positional encoding.

\subsubsection{Encoder}
\label{sec:encoder}
In this work we used the encoder inspired on transformer encoder proposed by \cite{dosovitskiy2020image} to capture global patterns and long-range dependencies within the time series light curves beyond the short-term correlations identified by the convolutional embedding. The Transformer encoder allows the model to weigh the importance of different time steps, enabling it to identify significant features relevant to exoplanet transit signals within the entire light curve, even amid the variability of the stars. \par 

The encoder uses the self-attention mechanisms to capture the relationships between different time steps within the light curves. This self-attention mechanism assigns varying levels of importance to different points in the sequence \citep{vaswani2017attention}. Thus, the model is able to focus on specific sections of the light curve, such as potential exoplanet transit signals, while disregarding irrelevant noise or variability star. The use of MSA layers facilitates the MSA mechanism in the encoder. This allows the attention mechanism, through multiple heads, to learn diverse and complementary representations of the light curve time series data, capturing different interactions and patterns that a single-head attention approach might miss. For this work, we employ $l=4$ self-attention layers and four heads. Additional experiments with different layer configurations and their impact on performance are detailed in Section \ref{sec:results_nn}.\par

The embeddings are initialized combining convolutional features and positional encodings (see Equation \ref{eq:z_embedding}). For each layer encoder $l$, the process is described as follow:

\begin{equation}
\mathbf{Z'}_{l} = \mathrm{MSA}({\mathrm{Norm}(\mathbf{Z}_{l-1})}) + {\mathbf{Z}_{l-1}}, \quad \mathrm{for} \quad l=1,..., L
\end{equation}

where $\mathbf{Z}_{l-1}$ is the input to $l$-th encoder layer, $\mathrm{Norm}$ represents layer normalization, $L$ is the number of layers of the Transformer encoder, and MSA is the MSA operation (MSA, see Appendix \ref{sec:appendix_att_mechanism}). \par


The layer normalization, as well as the residual connections, are integral to the encoder to contribute to the stability and efficiency of the model during training \citep{wang2019learning}. These techniques help prevent issues like vanishing or exploding gradients \citep{ba2016using}. In addition, the layer ensures that the inputs to the attention mechanism are consistently scaled \citep{dosovitskiy2020image}, stabilizing the distribution of activations. Consequently, the attention mechanism focuses more effectively on different parts of the light curve data, which includes complex patterns such as stellar variability. This stabilization not only improves the ability of the model to manage and interpret these transit signals but also improves training stability and convergence speed \citep{wang2019learning}. \par

Once the self-attention mechanism processes the input embeddings to captures contextual relationships, each time step passes through a position-wise feed-forward network, which also includes a residual connection:

\begin{equation}
\mathbf{Z'}_{l} = \mathrm{MLP}({\mathrm{Norm}(\mathbf{Z'}_{l})}) + \mathbf{Z'}_{l}, \quad \mathrm{for} \quad l=1,..., L
\end{equation}

where $\mathrm{MLP}$ is applied to the normalized output $\mathbf{Z'}_l$. This component also applies a non-linear transformation, specifically we use the Gaussian Error Linear Unit (GELU) \citep{hendrycks2016gaussian} activation, as proposed in the encoder of \citep{dosovitskiy2020image}. GELU is known for its smoother, probabilistic non-linearity, which facilitates more stable and efficient gradient flow during training. This enables the network to model complex and non-linear relationships within the light curve. Without this non-linearity, multiple layers would collapse into a single linear transformation. As the number of layers increases, the non-linear transformations allow the network to build progressively more complex representations \citep{goodfellow2016deep, dosovitskiy2020image}. \par

After $L$ stacked encoder layers, the final representation is obtained through layer normalization applied to the output of the last layer:

\begin{equation}
    \mathbf{Y} = \mathrm{Norm}(\mathbf{Z}_L)
\end{equation}

At this point, $\mathbf{Y}$ represents the normalized output of the linear encoder layer, which encodes the sequential information after the self-attention and feed-forward operations. 

Finally, we apply average pooling to downsample the features \citep{lecun2015deep}, defined as: 

\begin{equation}
\mathbf{Z}_{\mathrm{final}} = \frac{1}{T} \sum_{t=1}^{T} \mathbf{Z}_t
\end{equation}

where $T$ is the sequence length and $\mathbf{Z}_t$ represents the output features at each time step. Each $\mathbf{Z}_t$ is derived from the output $\mathbf{Y}$ after the normalization layer. The downsampling operation helps the NN to reduce the number of features in the subsequent layers, allowing them to focus on more abstract and comprehensive representations of the light curves. \par

\subsubsection{Final linear and sigmoid layer}
\label{sec:finallayer_softmax}
In the final layer, the feature vector passes through a final linear transformation. This fully connected linear layer takes the encoded representation from the transformer and projects it into a logits vector, which refers to the raw, untransformed values output by the model that represents the relative likelihood of each class. For binary classification, we apply a sigmoid activation function to the output of this linear layer, which converts the logits into a probability score \citep{goodfellow2016deep}. The sigmoid function ensures that the output is a value between 0 and 1, which represents the likelihood of the input belonging to the positive class, where a threshold is applied to this score to determine the final classification. The final transformation is defined as: \par 

\begin{equation}
    \hat{y} = \sigma (\mathbf{W} \cdot \mathbf{Z}_{\mathrm{final}} + b)
\end{equation}

where $\hat{y}$ is the predicted probability score, $\mathbf{W}$ is the weight matrix of the final layer, $\mathbf{Z}_{\mathrm{final}}$ is the output of pooled feature vector from the encoder, and $b$ is a bias term. The function $\sigma (x)$ represents the sigmoid activation function.



\subsection{Experimental setup}
\label{sec:experimental_setup}

Our model, detailed in this work, was developed using PyTorch \cite{paszke2019pytorch}, an open-source framework. Experiments were performed on an NVIDIA GeForce 1080 Ti with 11 GB of memory. The code is available online\footnote{\url{https://github.com/helemysm/FII_TransformerNN}}.


\subsubsection{Regularization \& Optimizers}
\label{sec:regularization}
Dropout is a regularization technique in which a subset of neurons is randomly deactivated during the training phase to prevent overfitting \citep{srivastava2014dropout}. Following the approach of \citet{vaswani2017attention}, we applied a dropout rate of 0.1 to the output of each layer. \par

We use the Adam optimizer, a variant of stochastic gradient descent (SGD) \citep{kingma2014adam}, because it adapts the learning rate dynamically as the network learns, which can lead to more efficient and effective training compared to using a fixed learning rate. To initialize the model, we set a relatively high learning rate of $\alpha = 0.001$, which progressively decreases as the network learns and the loss function value reduces. To implement this, we introduce a learning rate schedule, which lowers the learning rate over time \citep{goodfellow2016deep}. Starting with an initial learning rate of $\alpha = 0.001$, the rate is reduced by 20\% if there is no improvement in model performance after 5 to 10 epochs. This gradual decrease helps the model fine-tune its weights and improve convergence during training. In practice, this adjustment allows larger updates early in training and smaller adjustments later as the model improves its predictions \citep{goodfellow2016deep}. \par


\subsubsection{Training details}

We trained our model using the labelled data sets described in Section~\ref{sec:ground_truth}. Our model has planet and non-planet labels (1,0) as output options. During the training, we perform the data augmentation process described in Section~\ref{sec:augmentation}. Throughout each iteration of the training process, the model trains on diverse samples, which is made possible by the data augmentation techniques we have implemented. This ensures that the model does not see the same samples repeatedly, which helps enhance its generalization capabilities by exposing it to a variety of inputs. Additionally, during the training process, we implement the light curve injection process described in Section~\ref{sec:ground_truth}, which introduces new samples randomly throughout the training. We use all samples in batches of 120, which allows the model to efficiently process and learn from the data while optimizing memory usage. Each epoch consists of 400 iterations, with a duration of approximately 15 minutes. Our experiments were performed over 200 epochs, resulting in a total training time of around 30 hours. \par

%% file: training_process.tex
\section{Evaluation}
\label{sec:model_analysis}

Our evaluation test set consists of light curves from confirmed exoplanet candidates and false positives, such as EBs, and non-transit light curves. Since targets may have been observed across multiple sectors, we ensured that none of the targets in the evaluation set have associated targets in other sectors included in the training or validation sets. This separation guarantees that the model is tested not only on unseen light curves but also on unseen targets, preserving the integrity of the evaluation process. \par





\subsection{Performance evaluation metrics}
\label{sec:performancemetrics}


We evaluate our model using two classes: planet (positive instances) and not planet (negative instances). As we mentioned before, most of our dataset belongs to negative instances, and due to this imbalance, accuracy is not an appropriate metric as it can introduce bias by favouring the majority class. Therefore, we evaluate the performance of our model with metrics for unbalanced datasets and for binary classification. We focused on the Area Under the Receiver Operation Characteristic Curve (AUC-ROC) score metric and F1 score \citep{powers2020evaluation}. The AUC score is used for binary classification, and it provides a single value to summarize the true positive rate and the false positive rate. F1 score is the harmonic mean of the precision (how many identified exoplanet transits are correct) and recall (how many real exoplanet transits were detected).  \par

In our work, the AUC score helps us to evaluate the ability of the model to distinguish between true exoplanet transits and false positive signals like EB or non-transit signals. While the AUC score focuses on the model's overall classification performance, the F1 score provides an evaluation that considers both precision and recall. This helps us know how well the model identifies exoplanet transit while minimizing false positives and missed true transits. For each evaluation in all of the experiments, we perform 10-fold stratified cross-validation, taking the mean as the final result. \par



\subsection{Results}
\label{sec:results_nn}
Our model was evaluated with samples, which the network has not seen up to this point, which includes light curves from confirmed and known planets, EBs, and non-transit such as IS and V. This evaluation is based on metrics described in Section~\ref{sec:performancemetrics}. In addition, since the majority of light curves are expected to have no planetary transit signals, we evaluated the generalization capability of our model across different sectors of TESS by calculating the percentage of candidates detected by our implemented models. For this evaluation, we randomly selected a sample of light curves from the SPOC dataset and made predictions to determine the percentage containing exoplanet transit signals. As a result, our model identified that approximately 0.1\% of the light curves contain planetary transit signals, demonstrating its generalization ability across multiple sectors and its effectiveness in handling complex stellar variability. The main goal of this work is to improve the exoplanet candidate identification process, not only by increasing the likelihood of detecting potential candidates but also by reducing the false positive rate. This will improve the examination process for human vetting, as there will be fewer signals to analyze, with a higher probability that any given signal will contain an exoplanet. Currently, many DL approaches focus on the examination of pre-selected candidates \citep{valizadegan2022exominer}, as this remains a challenge in the field of exoplanet detection. \par

Table~\ref{tab:summary_results} shows the performance results for our model across various experimental configurations, including the use of centroid time series and background time series as input for our model. As a result, we found that incorporating the background time series significantly improves the performance of our model. In comparison, the centroids also contribute to performance improvement, but their impact is slightly less substantial. \par

Regarding the components of the attention mechanism, our default model is trained with four heads, as increasing the number of heads beyond this threshold yielded no significant improvement in performance while also leading to increased computational cost. We varied both the number of layers and the number of dimension embedding. Reducing $l = 2$ results in a 2 percent decrease in the AUC-ROC score, while setting $l = 8$ yields an AUC-ROC score equivalent to that of $l = 4$ and an F1 score with a 1 percent improvement for $l=8$. However, we opted to use $l=4$ as our default model as the reduced computational costs outweighed the increase in predictive performance for our use case. 

Additionally, the Table~\ref{tab:summary_results} highlights the impact of training the model without injected light curves, providing insights into how these injections affect the overall performance. Our results showed an improvement in the AUC-ROC score, and beyond that, we also made predictions to identify new candidates within a randomly selected sample from the SPOC dataset. Consequently, the model without data injection identified that approximately 0.36\% of the light curves contain planetary transit signals. This suggests that the model trained with data injection effectively reduces the number of light curves that need to be examined to determine potential planet candidates. These results suggest that training with injected data and the inclusion of background and centroid time series not only improves the performance but also allows for a more efficient evaluation of light curves. \par

\begin{table}
	\centering
	\vspace{5mm}
	\caption{Summary of the results in terms of AUC-score and F1-score. Default model is: $l = 4$, $d_{emb} = 512$, with centroid and background time series, and with injeted LCs. }
	\label{tab:summary_results}
	\vspace{4mm}
	\begin{tabular}{lcccc} 
		\hline
		Model &  Avg. auc score & Avg. F1-score \\
		\hline
        Default & 0.88 & 0.82  \\
        $l = 2$ & 0.84 & 0.80 \\
        $l = 8$ & 0.88 & 0.83 \\
        Without centroid & 0.87 & 0.80  \\
        Without background & 0.85 & 0.80  \\
        \textnormal{$d_{\textnormal{emb}}$} = 128 & 0.80 & 0.76\\
        \textnormal{$d_{\textnormal{emb}}$} = 256 & 0.84 & 0.79 \\
        Without injected LCs & 0.86 & 0.82 \\
        \hline		
	\end{tabular}
	\vspace{3mm}
\end{table}

%% file: catalogue.tex
\section{Searching New Candidates}
\label{sec:search_candidates}
In this section, we describe the process of identifying new exoplanet candidates using our trained model. We made predictions on SPOC data from sectors 1 to 26 with the aim of discovering previously undetected exoplanet candidates. Given that these sectors have already been extensively analyzed by numerous candidate search pipelines, employing both traditional methods and machine learning approaches, many possible candidates have already been catalogued. We filtered out all candidates that were already listed in the ExoFOP catalogue, as it is considered the authoritative source for previously identified candidates.
\par

\begin{figure}
\begin{center}
\vspace{5mm}
\includegraphics[width=0.48\textwidth]{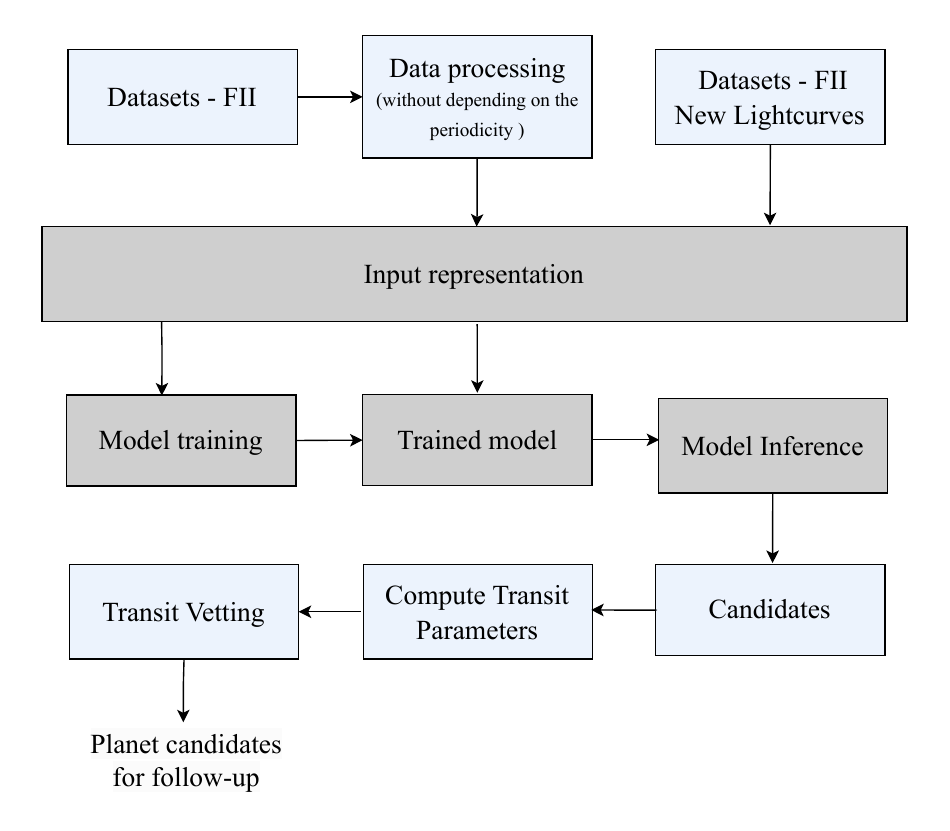}
\vspace{-2mm} 
\caption{\label{fig:pipeline}
General diagram for the identification of new exoplanet candidates based on model predictions. The blue boxes represent the inputs to the NN and the outputs after inference (candidates and transit vetting). The grey boxes are related to the NN, which includes the creation of the input representation, used to train the model and subsequently perform inference to find candidates. After inference, the resulting candidates are evaluated through transit vetting to identify potential exoplanets for follow-up observations.}
\vspace{-1mm}
\end{center}
\end{figure}

We developed a pipeline for the identification process, beginning with the inference of light curves to extract potential candidates. Our model evaluated a total of 4.1 million light curves, identifying 0.1\% with a probability of containing planetary transit signals. Among the light curves identified, both single-planet multi-transit and single-transit events were detected, each requiring a specific vetting approach. Figure~\ref{fig:pipeline} shows the general procedure for identifying new candidates. Once our trained model performs inference on new light curves and identifies those that contain potential planetary transits, we calculate the transit parameters for these candidates. We used various techniques for vetting the light curves identified. For detailed information on these examinations, refer to Section \ref{sec:periodic_candidates}, Section \ref{sec:single_candidates} and Section \ref{sec:multiplanet_candidates}. \par

\subsection{Single-planet multi-transit light curve candidates}
\label{sec:periodic_candidates}
Single-planet multi-transit light curve candidates are those where multiple transit events are observed at regular intervals in the light curve, indicating a consistent orbital period around the host star. Some planetary systems may not show multiple transits due to observational time, such as long orbital periods that exceed the duration of the observations. Additionally, systems with TTVs can cause variations in the timing of transits, making them deviate from strict periodicity. The multiple transits in the light curve at regular intervals not only help confirm the presence of the exoplanet candidate but also enable more precise estimations of its orbital parameters. \par

To verify these candidates, we first apply the Box-Least Squares 
\citep[BLS;][]{kovacs2002box} periodogram to compute the transit parameters. BLS is an algorithm designed to search for periodic signals in light curves, specifically focusing on those that have a box-shaped transit shape. After computing the transit parameters, we calculate the radius of the planet $r_{\mathrm{p}} = r_{\ast} \sqrt{d(c+1)}$, using the radius of the target star $r_{\ast}$ and the target's background contamination $c$. The stellar radius $r_{\ast}$ and the $c$ are obtained from the TIC, accessed through the Mikulski Archive for Space Telescopes (MAST) using \texttt{Astroquery} package. When the stellar radius is not available in the TIC, we use the  Gaia mission's \citep{vallenari2023gaia} data release 3 as a secondary source. We discard planet candidates with a predicted radius greater than 1.8$R_{\mathrm{Jupiter}}$. This threshold is set to focus on identifying planets that lie within the typical size range of gas giants and smaller exoplanets, as larger objects are more likely to be low-mass stars, brown dwarfs, or other non-planetary astrophysical phenomena \citep{hodvzic2018wasp, carmichael2020two}. According to the BLS outputs, approximately $20\%$ of the signals have physical parameters consistent with a planetary transit. \par


After using BLS, we used the Discovery and Vetting of Exoplanets (DAVE) tool, developed by \cite{kostov2019discovery}. Only those signals that passed the initial BLS filter proceeded to this subsequent examination. DAVE evaluates the candidates by examining several factors, among them checking for secondary transit signals, evaluating changes in photometric centroid positions during and outside of the transit events, and comparing odd against even transits. The check for secondary transit signals helps identify binary star systems that might mimic planetary transits, as these typically exhibit a detectable secondary eclipse. The examination of photometric centroid shifts ensures that the detected signal is originated from the target star and not from nearby stars. In some cases, DAVE classified a TIC as a candidate in one sector but flagged it as a false positive in another. In these instances, we consider the TIC to be a false positive. After human vetting and reviewing the output of DAVE for candidates with a positive disposition, $\sim 25\%$ of the cases correspond to valid candidates. \par

\begin{figure}
\begin{center}
\vspace{5mm}
\includegraphics[width=0.48\textwidth]{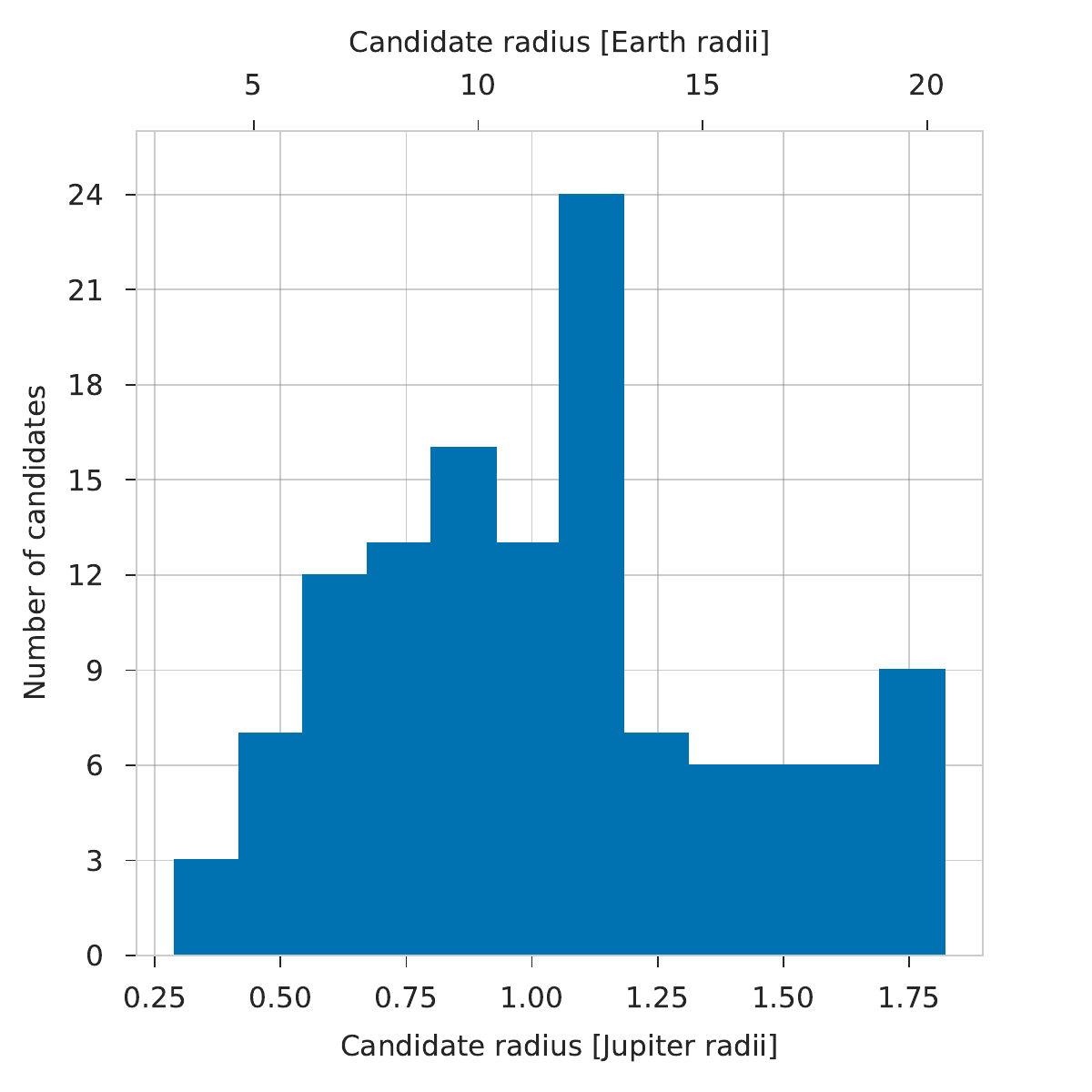}
\vspace{-2mm} 
\caption{\label{fig:periodic_transiters_radii}
Distribution of single-planet multi-transit light curve candidate radii measured in Jupiter raidii.}
\vspace{-1mm}
\end{center}
\end{figure}

\begin{figure}
\begin{center}
\vspace{5mm}
\includegraphics[width=0.48\textwidth]{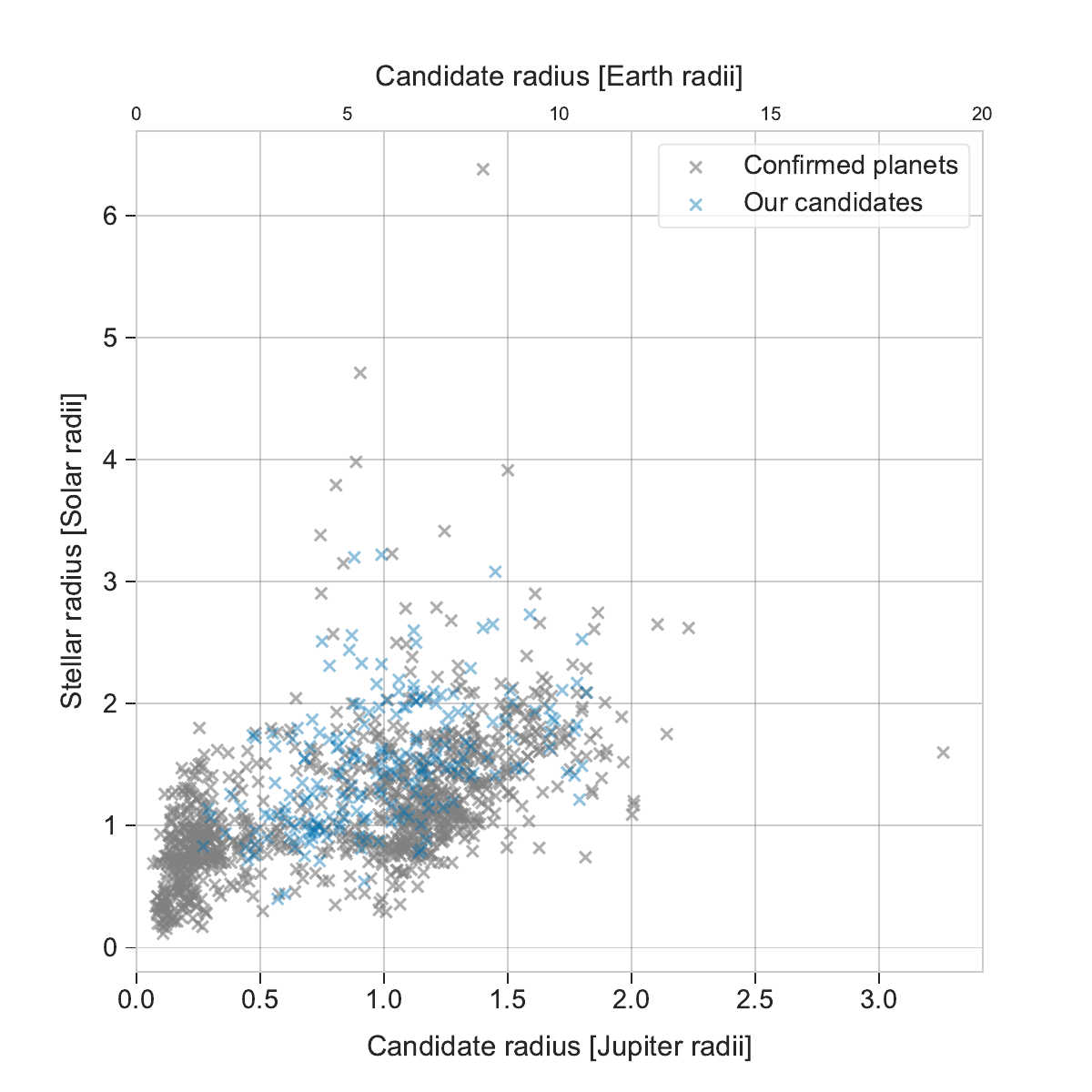}
\vspace{-2mm} 
\caption{\label{fig:targetr_radii}
Relationship between radius candidate and stellar radius. Each point represents a confirmed planet from ExoFOP-TESS or our candidates.}
\vspace{-1mm}
\end{center}
\end{figure}

The results from our model predictions, after completing the vetting process, are summarized in Table~\ref{tab:periodictransit_candidates}. The table presents the candidates along with their respective transit parameters, including target magnitude (TMag), transit depth, transit epoch, orbital period, transit duration, and radius. These parameters provide a detailed overview of the characteristics of the candidate. Figure~\ref{fig:periodic_transiters_radii} shows the distribution of radii for the identified candidates measured in Jupiter radii, providing insight into the range and frequency of detected planet sizes. The majority of the candidates are concentrated within a radius range between 0.75 and 1.25 Jupiter radii, with a significant peak around 1 Jupiter radius. We did not detect any Earth-sized exoplanets or smaller candidates with radii below 0.3 Jupiter radii. Figure \ref{fig:targetr_radii} shows that stars with radii between 0 and 1 solar radii host a large number of confirmed planets with radii between 0 and 2 Earth radii (upper x-axis), suggesting that smaller planets are more commonly found around smaller stars.  In contrast, our detected candidates (blue points) are concentrated in the 0.75–1.25 Jupiter radii range and mostly around stars between 0.8 and 2 solar radii. The absence of smaller planets in our findings can be attributed to several factors. For example, smaller planets, such as those near the size of Earth, typically produce shallower transits. Traditional methods such as BLS depend heavily on periodicity and may be better to detect shallower transit signals, since it considers multiple potential periods, thereby increasing detection likelihood. Our NN still has challenges in identifying these signals, especially in the presence of noise or stellar variability. Unlike BLS, our approach does not depend on periodicity, which could explain why our NN detects smaller candidates less frequently. \par

\begin{figure}
\begin{center}
\vspace{5mm}
\includegraphics[width=0.48\textwidth]{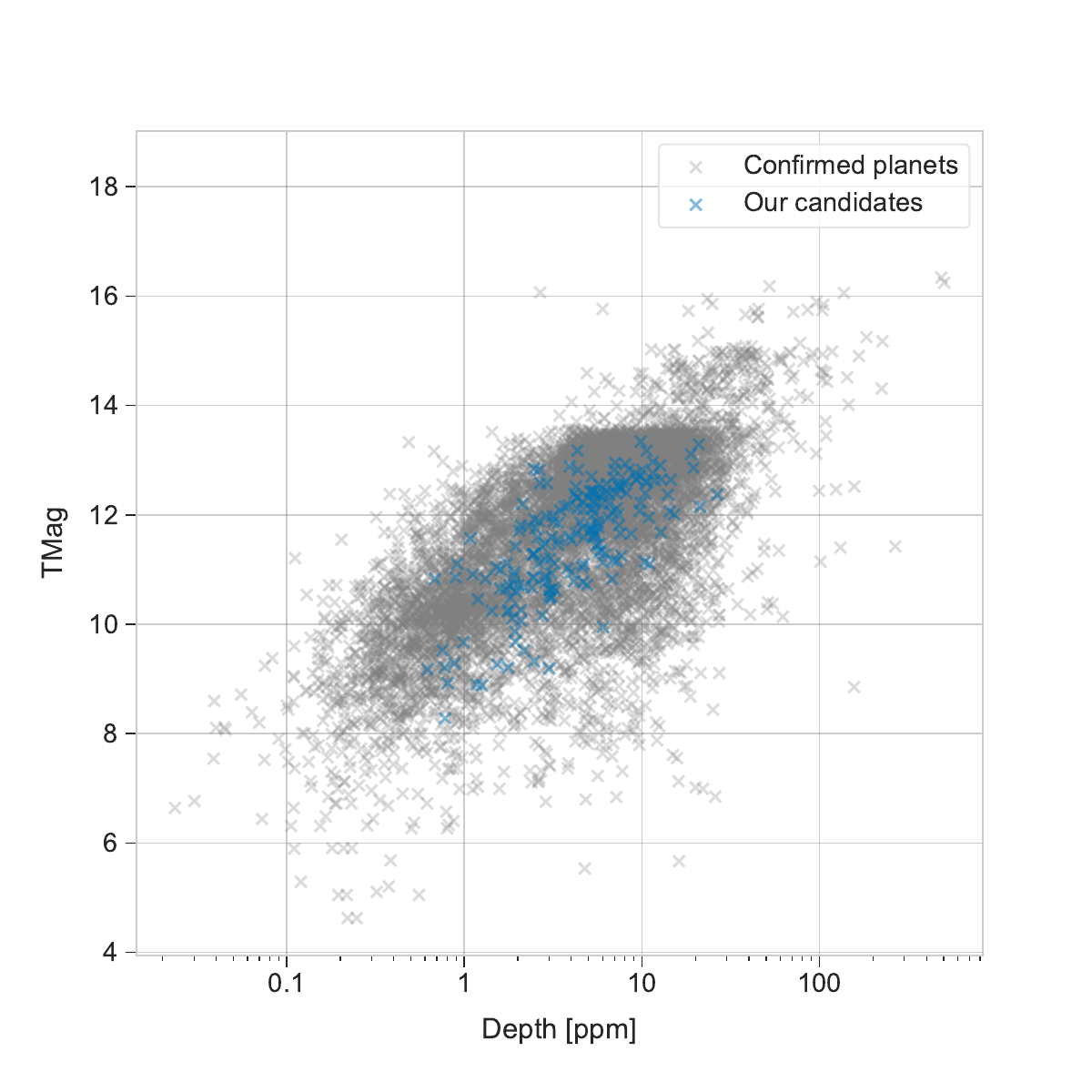}
\vspace{-2mm} 
\caption{\label{fig:tmag_vs_depth}
Relationship between TESS magnitude and transit depth. Each point represents a confirmed planet from ExoFOP-TESS or our candidates.}
\vspace{-1mm}
\end{center}
\end{figure}

We compared our exoplanet candidates with ExoFOP-TESS-confirmed planets. Figure~\ref{fig:tmag_vs_depth} shows the relationship between TESS magnitude and transit depth for confirmed exoplanets and for our candidates, indicating that our candidates are concentrated in an intermediate region of depth and magnitude compared to the confirmed planets. The relationship between the radius of exoplanets measured in Jupiter radii and their corresponding orbital periods is illustrated in Figure~\ref{fig:radius_vs_orbital_period}. The majority of confirmed exoplanets have orbital periods of less than $\sim$ 5 days, but there are also many with longer periods, showing great diversity. Among our candidates showing multi-transits, we also find some with periods shorter than 5 days, extending up to 25 days. \par


We identified two candidates with radii slightly exceeding the threshold of 1.8 $R_{\mathrm{Jupiter}}$. For candidates that exceed this threshold, we calculated their equilibrium temperature of the planet $T_{\mathrm{eq}}$ using Equation 4 from \cite{mendez2017equilibrium}, which is calculated from the balance between the stellar flux received from the host star and the flux absorbed by the planet \citep{selsis2007habitable}. For candidate TIC 385921743, we calculated an equilibrium temperature of $T_{\mathrm{eq}}$ = 2449.11 $\mathrm{K}$, and a radius of 1.85 $R_{\mathrm{Jupiter}}$. The radial velocity error (RV error) obtained from Gaia is 0.91 $\text{km s}^{-1}$. The second candidate, TIC 275614831, has an $T_{\mathrm{eq}}$ of 3870.08 $\mathrm{K}$, and a calculated radius of 1.82 $R_{\mathrm{Jupiter}}$, and an RV error of 1.50 $\text{km s}^{-1}$. 
Despite being slightly above the threshold, these candidates were included for further consideration due to their close proximity to the limit. The $T_{\mathrm{eq}}$ for both candidates are near the upper range observed for known hot Jupiters but remain within it, suggesting they could be gas giants in tight orbits around their host stars, making them interesting targets for additional follow-up observations. Moreover, the radial velocity errors are relatively low, which suggests a reasonable level of confidence in their classification as planetary candidates.


\subsection{Single-transit light curve candidates}
\label{sec:single_candidates}
Single transiters are exoplanet candidates where only one transit has been observed in the light curve of a single TESS sector. This typically happens because TESS observes each sector for 27 days, and the orbital period of these candidates is longer than the duration of a single sector light curve. Additional transits may appear in other sectors depending on the orbital period and the timing of observations. Longer-period transiting exoplanets are particularly valuable as they enable us to measure the densities of warm and cool planets \citep{wang2021aligned}. They often present unique opportunities for discovery, as they can represent long-period planets that may not be detectable through traditional methods, such as BLS. Furthermore, once these transits are detected, they will require further ground-based follow-up observations to determine the orbital period and differentiate from potential astrophysical or statistical false positives \citep{sullivan2015transiting}. Detecting and validating these signals is more challenging compared to multi-transiters, which show multiple transits within the light curve. \par

From the NN predictions, we perform a visual inspection and identify those that have only one transit signal corresponding to $\sim$ 400 TIC IDs, with the remaining cases evaluated as multi-transiters (see Section \ref{sec:periodic_candidates}). We verified that the detected transit signal originates from the target star itself and not from nearby stars that could contaminate the observation. Next, we discard those TIC IDs that have an RV error $>$ 2 $\text{km s}^{-1}$, with the RV measurements obtained from Gaia. This criterion allows us to eliminate potential false positives resulting from measurement uncertainties in radial velocity data. This filter was consistent with the radius of the candidates, as those with an RV error $>$ 2 $\text{km s}^{-1}$ generally had radii larger than 1.8 $R_{\mathrm{Jupiter}}$, with the exception of TIC 438122862, which has a radius of 0.93 $R_{\mathrm{Jupiter}}$ and presents only one single transit in Sector 6. After applying these filters, we verify whether the transits were detected in sectors beyond 26 to ensure consistency across multiple observations of the same transit event. Finally, for the TIC IDs identified, we extend the review to sectors from subsequent sectors, confirming the persistence of the signals or identifying any new ones. As a result, $\sim$ 20\% passed the vetting process, leaving 82 validated as single transiters.

\begin{figure}
\begin{center}
\vspace{5mm}
\includegraphics[width=0.48\textwidth]{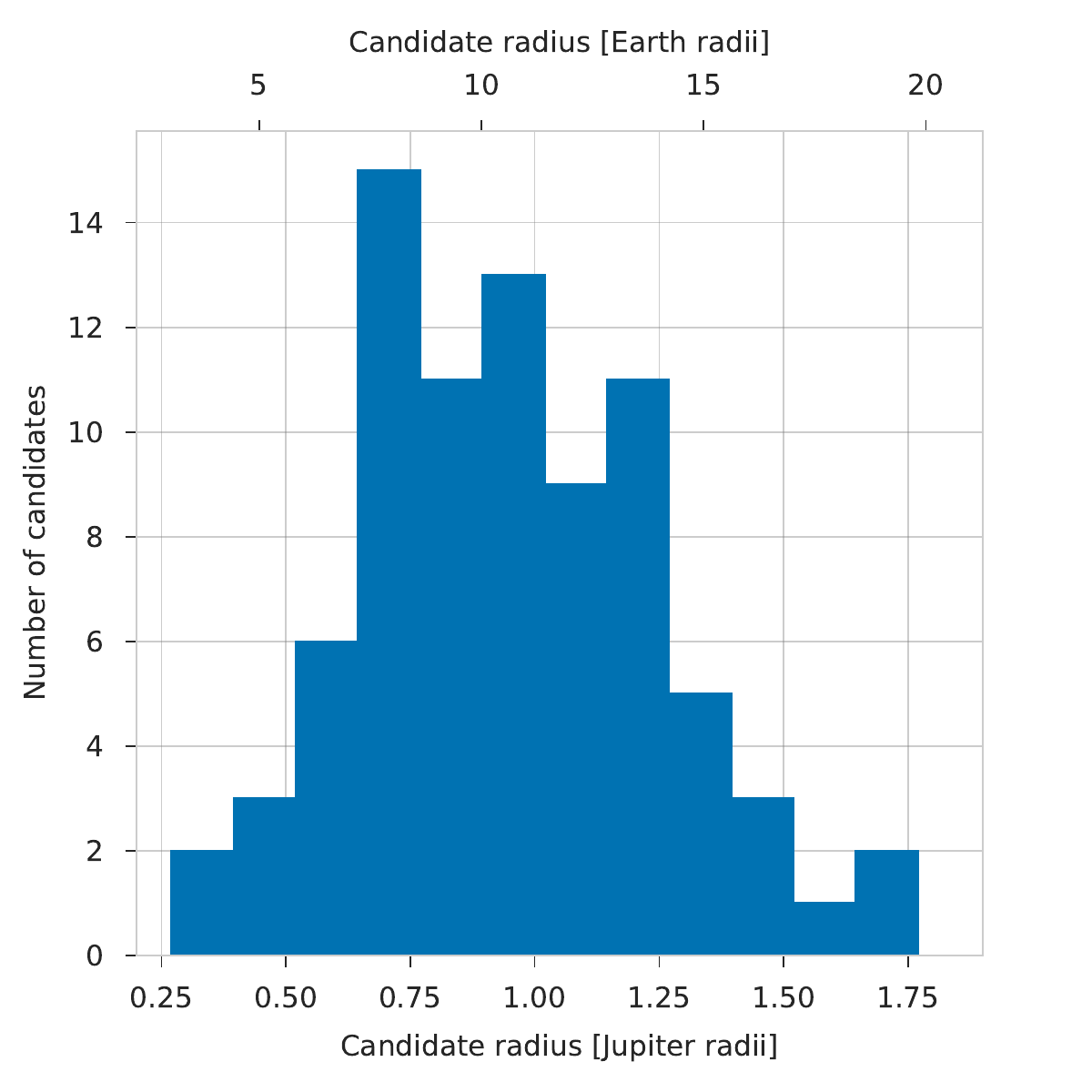}
\vspace{-2mm} 
\caption{\label{fig:single_transiters_radii}
Distribution of single-transit light curve candidate radii, measured in Jupiter radii.}
\vspace{-1mm}
\end{center}
\end{figure}

Following our model predictions and the subsequent vetting process, we present in Table~\ref{tab:singletransit_candidates} a detailed list of candidates identified as single transiters. The distribution of single transit candidates as a function of their radii is shown in Figure~\ref{fig:single_transiters_radii} where most candidates have radii between 0.5 and 1.0 Jupiter radii, with a peak around 0.75 Jupiter radii (approximately 8.5 Earth radii). 

\begin{figure*}
\begin{center}
\vspace{5mm}
\subfloat{\includegraphics[width=0.93\textwidth]{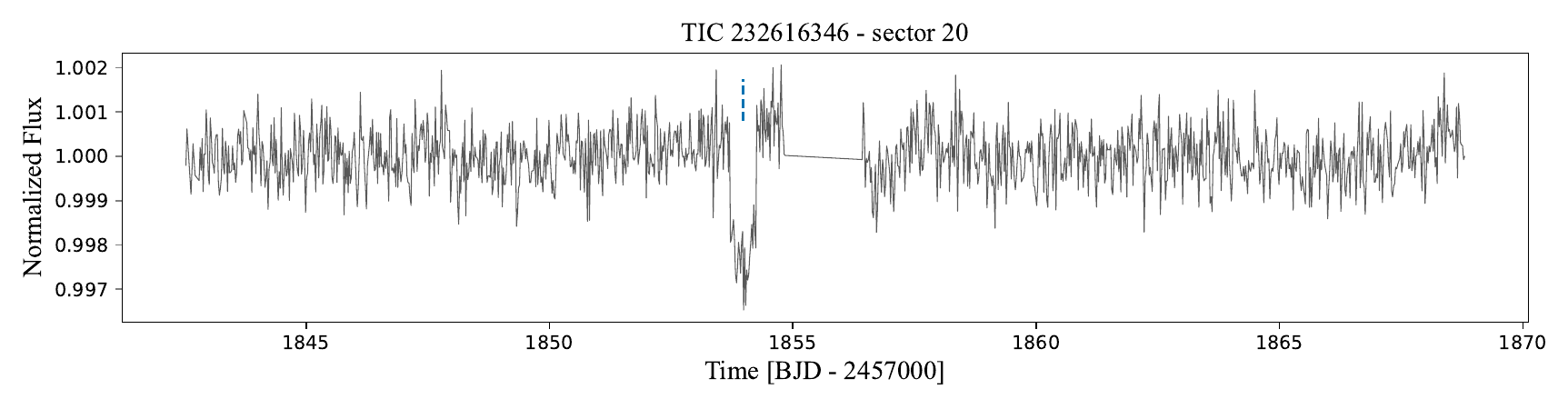}\label{fig:232616346_s20_lc}}\\
\vspace{2mm}
\subfloat{\includegraphics[width=0.93\textwidth]{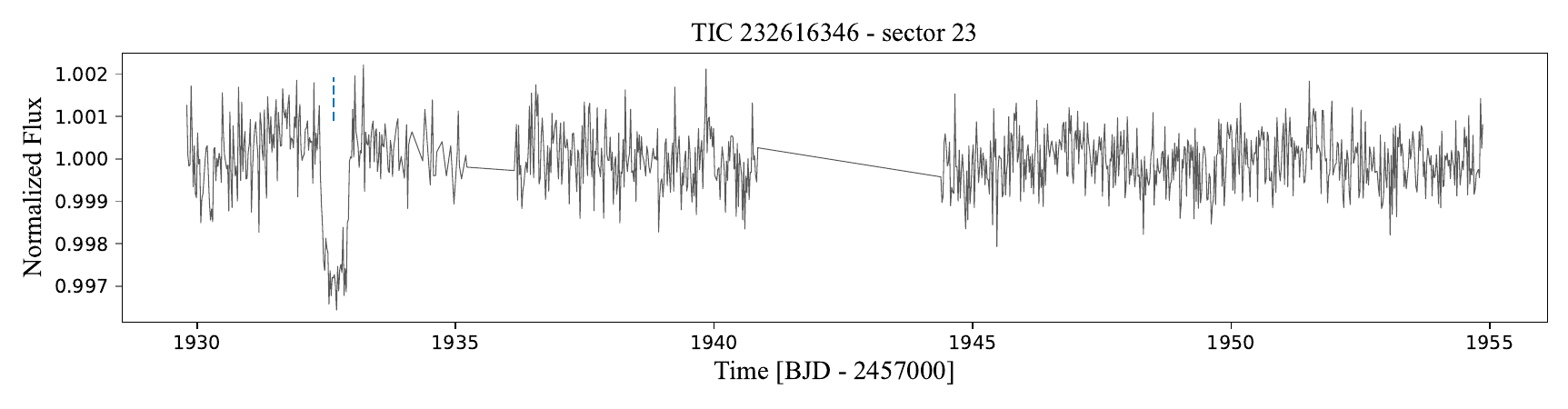}\label{fig:232616346_s23_lc}}
\vspace{2mm} 
\caption{Light curves for TIC 232616346 in sectors 20 and 23, showing the same transit event with consistent depth, validating the detection of the same candidate in both sectors. \label{fig:232616346_lc}}
\vspace{-1mm}
\end{center}
\end{figure*}

\begin{figure*}
\begin{center}
\vspace{5mm}
\subfloat{\includegraphics[width=0.93\textwidth]{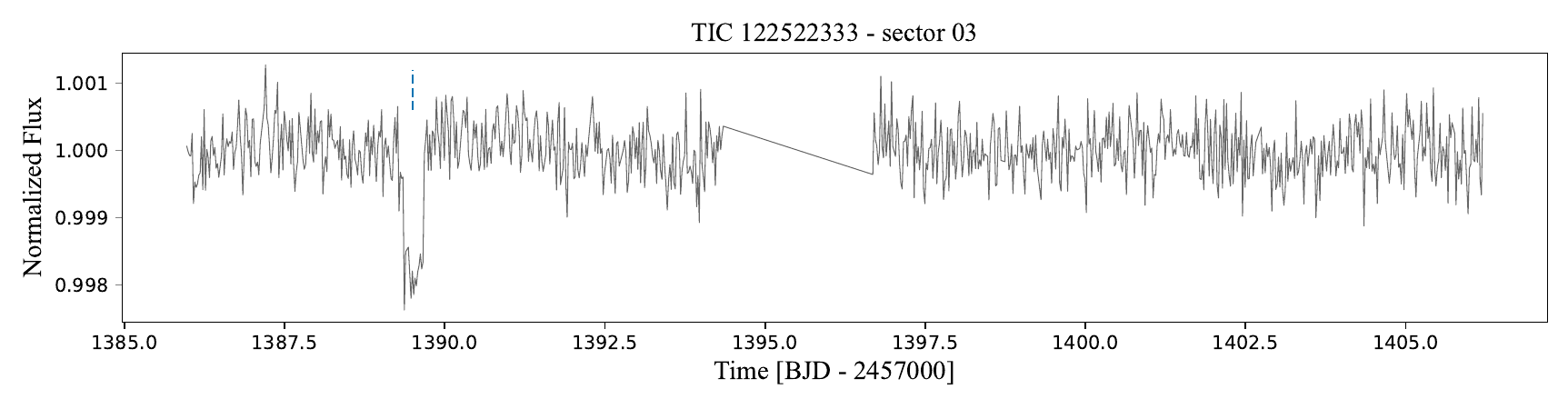}\label{fig:122522333_s03_lc}}\\
\vspace{2mm}
\subfloat{\includegraphics[width=0.93\textwidth]{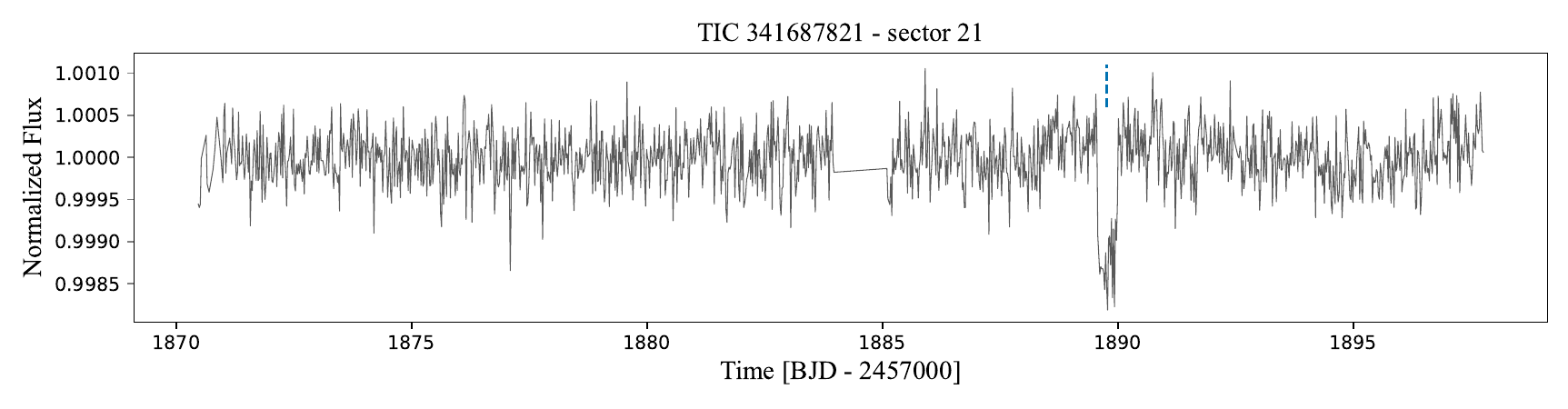}\label{fig:341687821_s21_lc}}
\vspace{2mm}
\subfloat{\includegraphics[width=0.93\textwidth]{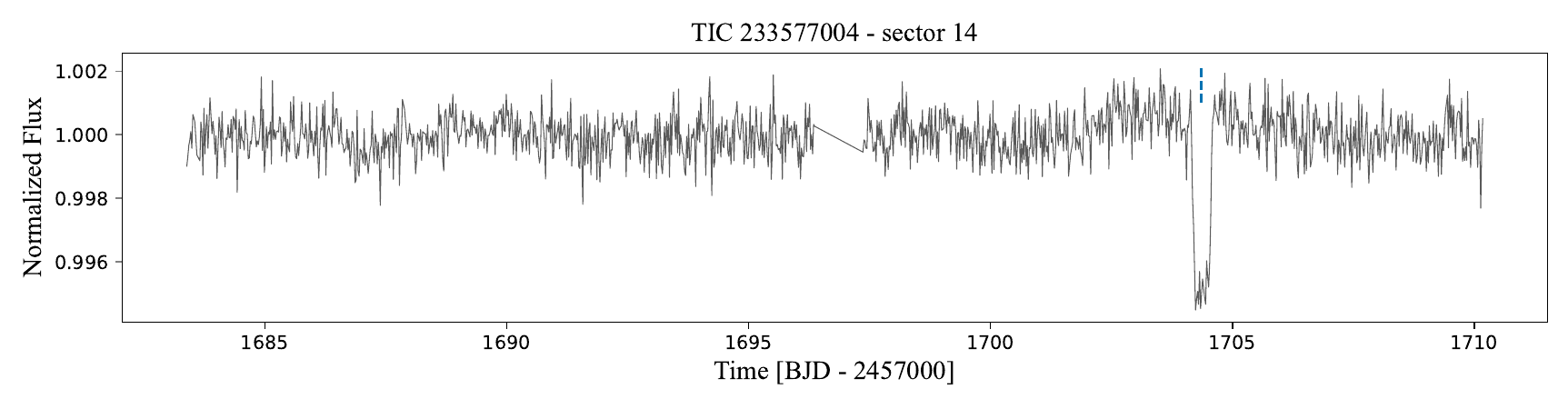}\label{fig:233577004_s14_lc}}
\vspace{2mm} 
\caption{
Three examples of single transiters from TESS observations for TIC 122522333, TIC 341687821 and TIC 233577004. Each candidate shows a single-transit event in the respective sector. While these candidates were also detected in other sectors (TIC 122522333 in sector 4, TIC 341687821 in sector 14, and TIC 233577004 in sectors 21), the figure shows only the transit events from the sectors presented.
\label{fig:single_transit_candidates}}
\vspace{-1mm}
\end{center}
\end{figure*}

An example of a candidate identified by our NN in two different sectors is TIC 232616346. Figure~\ref{fig:232616346_lc} shows the light curves of this candidate identified in sectors 20 and 23. Both sectors show the same transit event with the same depth, which validates that the candidate is the same in both sectors. Similarly, other candidates were also detected in multiple sectors, while some were identified in only a single sector because additional transits may have occurred when TESS was not observing that sector. Figure~\ref{fig:single_transit_candidates} presents three additional examples of single transit candidates: TIC 233577004 in sector 14, TIC 341687821 in sector 21, and TIC 122522333 in sector 3. Each of these candidates displays a clear single transit event, with some having detections in other sectors and others identified in only one sector. These light curves show the type of signals identified by our model, suggesting that these candidates may correspond to longer-period exoplanets. \par

Moreover, these examples underscore the challenges in analyzing single transit events. Our model demonstrates the ability to detect the presence of a transit signal, but it is essential to perform careful validation to discard false positives and ensure the reliability of the identified candidates. In this sense, continued monitoring and follow-up observations will be crucial for confirming these candidates.  \par

\subsection{Multi-planet system candidates}
\label{sec:multiplanet_candidates}
Multi-planet systems are planetary systems that host more than one planet orbiting the same star. These systems usually show multiple transits with different depths in their light curves. The varying depths can be attributed to the planets of different sizes from the host star. Additionally, the orbital period of the transits differs due to the different distances of the planets from the host star. 

\begin{figure*}
\begin{center}
\vspace{5mm}
\subfloat{\includegraphics[width=0.93\textwidth]{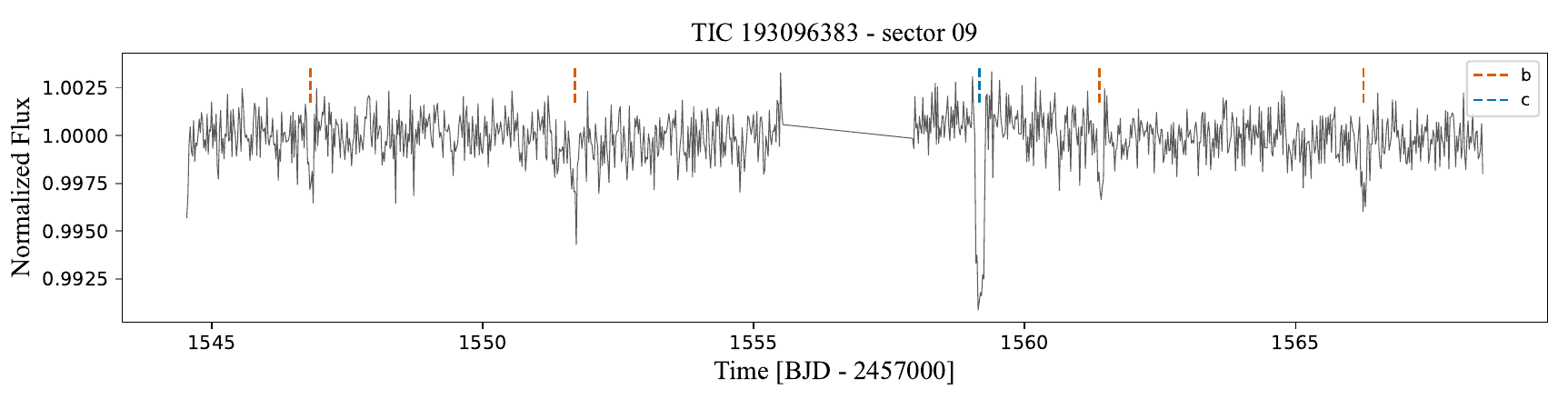}\label{fig:193096383_s09_lc}}\\
\vspace{2mm} 
\caption{
Light curve for TIC 193096383 in sector 9, showing the transits of candidates b (orange) and c (blue). The transits of candidate b have the same depth, which differs from the depth observed in the transit of candidate c, suggesting a potential multi-planet system.\label{fig:multiplanet_system}}
\vspace{-1mm}
\end{center}
\end{figure*}

Our NN is trained to identity transit signals in light curves, regardless of whether these signals have identical depths or orbital periods. The NN distinguishes between transit signals through the analysis of key features of the light curve, such as the shape of the dip, the variation in the brightness fo the star and the timing of the signal. As a result, the NN is able to identify multi-planet systems. Among these, TIC 193096383, TIC 221567884, TIC 118798035, and TIC 400049903 were detected across sectors 1 - 26. TIC 193096383 is illustrated in Figure~\ref{fig:multiplanet_system}, which shows the transits of two potential exoplanet candidates, ``b''(orange) and ``c' (blue). To validate these candidates, we implemented an approach that combines analyses of single-planet multi-transit and single-transit light curve candidates, as detailed in sections ~\ref{sec:periodic_candidates} and ~\ref{sec:single_candidates}. Although a detailed examination of multi-planet systems is beyond the scope of this work, we underscore the importance of further investigation of these TIC IDs. \par

\subsection{Significant candidates of special interest}

As a result of this work, certain candidates emerge as particularly remarkable in the search for exoplanets due to their unique characteristics or potential for further study. We identify these candidates while noting that a more detailed analysis of them is beyond the focus of this work. We highlight the need for continued investigation and validation to uncover insights into planetary formation and system dynamics. \par

\subsubsection{TIC 221567884}
Our model initially detected TIC 221567884  as a candidate in sector 12. After further analysis using BLS, we identified two transits with different depths, suggesting it could be a multi-planet system. To discard the possibility that the TIC 221567884 belongs to an EB, we verified additional sectors $>$ 26 for the same TIC 221567884 to ensure the consistency of the signals across multiple observations. We identified transits in sectors 39 and 66, with a particular possible mutual transit event in sector 66. In systems with more than one planet, there is a unique opportunity for photometric mutual events, where planets pass in front of or obscure each other as seen from Earth. A mutual event specifically refers to this overlap between two planets from our perspective on Earth \citep{ragozzine2010value}. If confirmed, this mutual event would be of particular interest. \par

\begin{figure*}
\begin{center}
\vspace{5mm}
\subfloat{\includegraphics[width=0.93\textwidth]{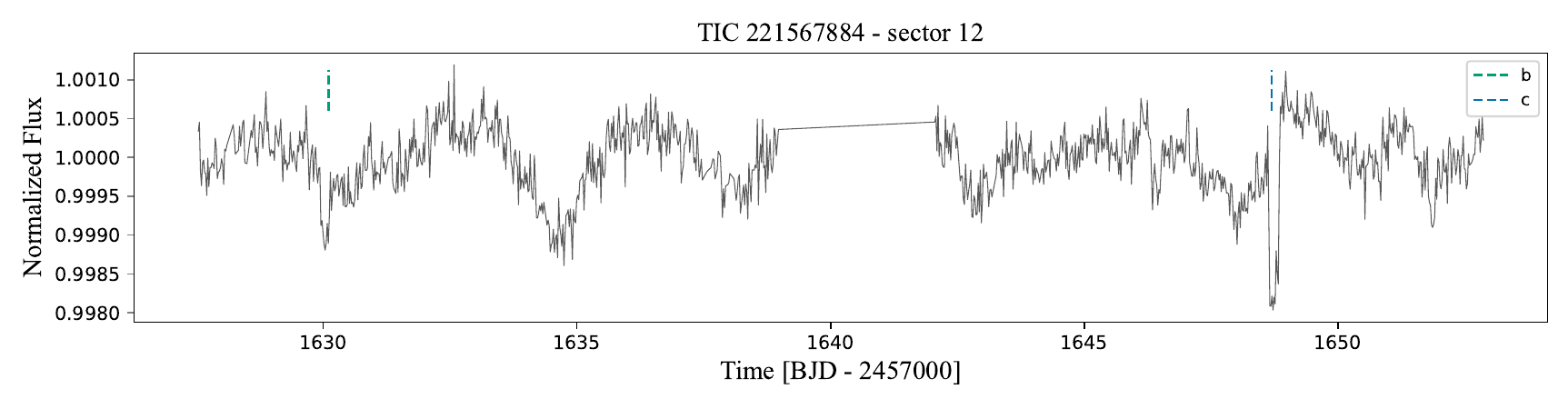}\label{fig:221567884_s12_lc}}\\
\vspace{2mm}
\subfloat{\includegraphics[width=0.93\textwidth]{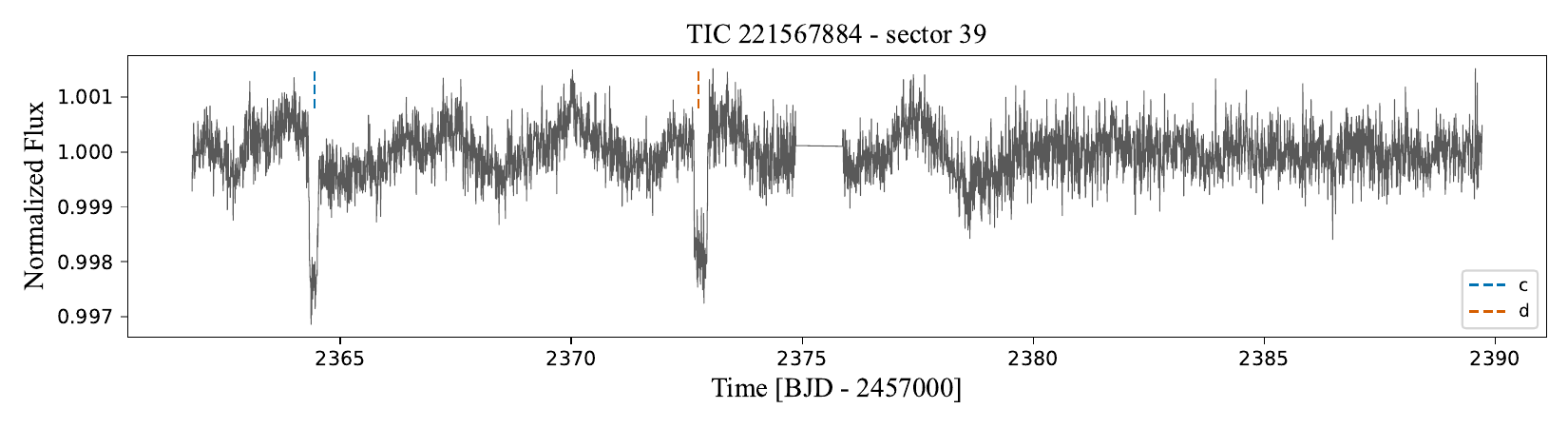}\label{fig:221567884_s39_lc}}\\
\vspace{2mm}
\subfloat{\includegraphics[width=0.93\textwidth]{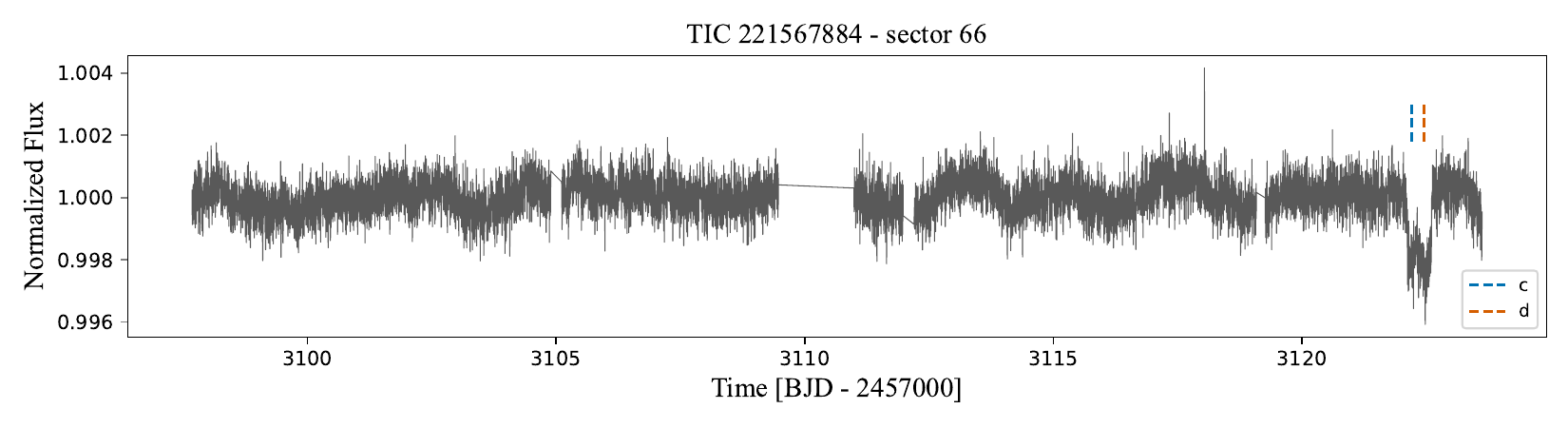}\label{fig:221567884_s66_lc}}\\
\vspace{2mm} 
\caption{
Lightcurves for TIC 221567884 across sectors 12, 39, and 66, illustrating transit events ``b'', ``c'' and ``d'', and a potential mutual transit event between candidates ``c'' and ``d'' in sector 66. \label{fig:221567884_s_lc}}
\vspace{-1mm}
\end{center}
\end{figure*}

Figure~\ref{fig:221567884_s_lc} shows light curves for the TIC 221567884 across three different sectors from the TESS mission (sectors 12, 39, and 66). In the top panel (sector 12), there are slight flux variations with two transit events marked as ``b'' (green) and ``c'' (blue); both transit events are described in Table~\ref{tab:multiplanet_candidates}. In the middle panel (sector 39), the transit of candidate ``c'' (blue) is observed again, showing the same depth as in sector 12, along with the transit of candidate `d'' (orange). Finally, the bottom panel (sector 66) highlights the previously mentioned mutual transit event of candidates ``c'' and ``d'', with depths consistent with those observed for candidate c in sectors 12 and 39 and for candidate d in sector 39. This consistency supports their identification as potential exoplanets and suggests a potential indication of planetary interactions within the system.  \par

\subsubsection{TIC 118798035}
TIC 118798035 is a significant candidate within a multi-planet system, showing noteworthy TTVs. TTVs indicate deviations from the expected transit times based on a basic Keplerian model, reflecting the gravitational interactions between the planets in the system. This suggests that exoplanets in multi-planet systems often follow non-Keplerian orbits, leading to variations in the timing and duration of their transits \citep{agol2017transit}. \par

\begin{figure*}
\begin{center}
\vspace{5mm}
\subfloat{\includegraphics[width=0.93\textwidth]{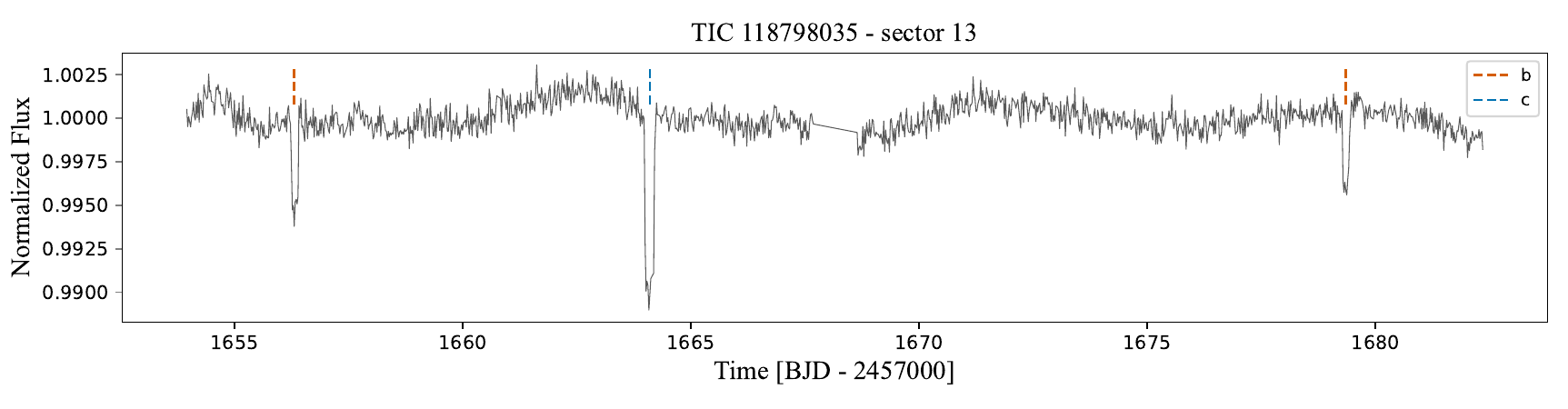}\label{fig:118798035_s13_lc}}\\
\vspace{2mm}
\subfloat{\includegraphics[width=0.93\textwidth]{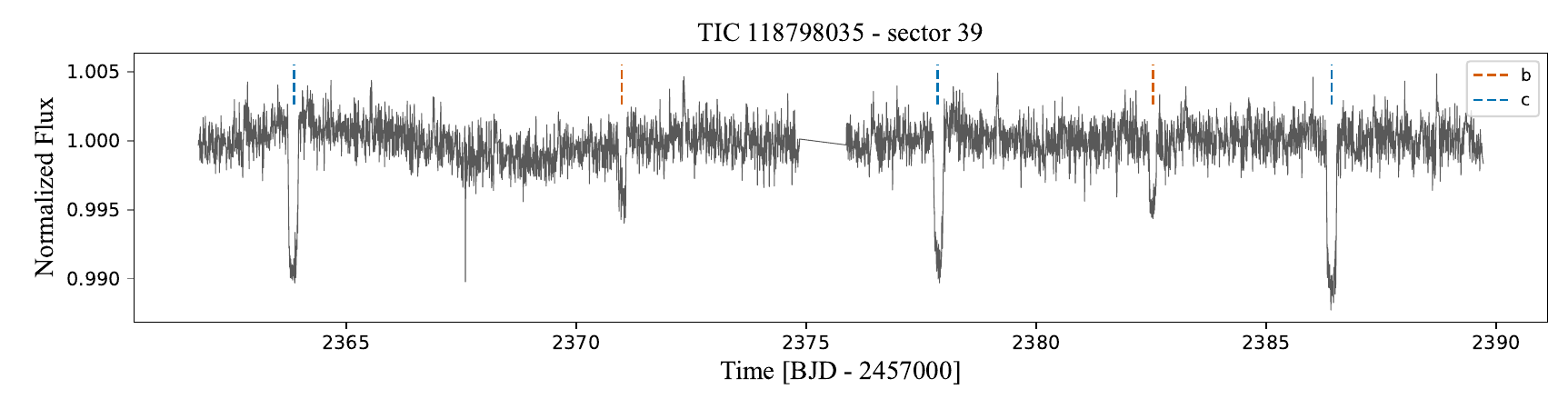}\label{fig:118798035_s39_lc}}
\vspace{2mm} 
\caption{Light curves for TIC 118798035 illustrating a system with two exoplanet candidates. In sector 13, transit signals for candidates ``b'' (orange line) and ``c'' (blue line) are detected. The depths are consistent across both sectors, with TTVs for candidate ``c'' in sector 39. \label{fig:118798035_lc}}
\vspace{-1mm}
\end{center}
\end{figure*}

TIC 118798035 was identified as a candidate by our model in sector 13. To further investigate this system, we extended our analysis to sectors beyond 26. In sector 39, additional transits from potential candidates were observed, providing further evidence of a multi-planet system with TTVs. Figure~\ref{fig:118798035_lc} presents the light curves for this possible multi-planet system, with the top panel showing the light curve from sector 13 and the bottom panel showing the light curve from sector 39. In sector 13, transits for candidates ``b'' and ``c'' are clearly visible. In sector 39, the transit of both candidates ``b'' and ``c'' appear again with consistent depths. Furthermore, TTVs for candidate ``c'' is showed in sector 39. \par

\subsection{Analysis of false positives}

Despite incorporating information on centroids and background time series to help the model learn to distinguish false positives, some false positives are still detected by our NN. An example of this is TIC 296945443 from TESS Sector 12, which was detected by our NN as an exoplanet candidate but was later identified as a false positive by DAVE. The analysis of DAVE showed a shift in the photometric centroids, which was consistent with background star contamination. \par

Our NN was trained on SPOC light curves, which are calculated using various photometric apertures. In the case of TIC 296945443, the aperture selected for Sector 12 did not capture the brightness variation of a nearby contaminating star, resulting in no detected centroid shift and leading to a false positive. We also examined light curves from other sectors for the same target, and in Sector 38, SPOC detected a centroid shift that confirmed the presence of a nearby object causing the transit-like signal. Our NN detected this target in Sector 38 as a candidate but with a lower probability compared to the high probability assigned in Sector 12. This analysis is consistent with the results presented in Section \ref{sec:results_nn}, demonstrating that incorporating centroid information does improve detection probability. However, it also highlights that further improvement can be made to reduce false positives. For example, one possible solution is to increase the training set, including more examples of false positives with centroid shifts. This additional data could help the model better recognize cases where background stars or centroid shifts cause false positives and, thus, improve the ability to distinguish true exoplanet candidates from false positives.


\begin{table}
    \centering 
    \caption{List of single-planet multi-transit light curve candidates.}
    \label{tab:periodictransit_candidates}
    \begin{tabular}{lccccc} 
        \hline 
        \hline
        TIC ID & Transit & Radius & Depth & Orbital& Duration \\ 
        & Epoch & (Jupiter &  & Period& \\
        &  & radii) & (ppm) & (days)& (hours)\\
        \hline 
        \begin{filecontents*}{candidates/periodictransitcandidatepart1.csv}
TICID,tdepth,radius,orbitalperiod,duration,epoch,TessMag,radiustarget,temperaturetarget,RA,DEC,RVerror
123365281,11518,1.24,2.606,2.52,1386.206,12.97,1.15,5918.0,31.2272,-30.9494,2.487
88507544,21209,1.14,2.154,1.79,1411.785,12.15,0.78,4737.0,47.7706,-29.7579,0.473
245076932,4623,0.98,21.605,4.81,1601.167,11.04,1.3,6123.0,198.0637,-51.2144,0.46
32421578,3182,0.68,4.988,2.88,1390.559,11.88,1.2,6196.0,27.1367,-33.6789,1.101
138733600,9198,1.08,3.339,2.88,1387.988,12.76,1.12,5463.0,33.1782,-38.9509,1.37
302753992,682,0.75,0.801,1.44,1325.620,10.83,2.51,6083.0,322.9834,-24.3044,0.473
39017649,8071,1.42,10.618,3.12,1332.992,12.93,1.56,5653.0,11.7008,-67.5199,2.296
29756364,5343,0.90,1.173,1.44,1325.815,12.42,1.23,5679.0,315.9598,-24.9421,1.771
341497604,2666,1.09,1.687,1.69,1325.815,12.58,1.97,6138.0,318.5934,-61.2951,1.064
206584800,2443,0.86,1.097,1.44,1355.234,11.28,1.74,6264.0,20.3209,-54.7649,0.624
70514245,19522,1.15,4.048,1.55,1387.838,12.86,0.8,4780.0,32.7182,-34.9084,1.187
27986137,10000,1.03,3.438,1.56,1389.127,12.82,1.03,5697.0,10.3337,-22.8677,1.749
164695417,3050,0.69,3.597,1.32,1387.019,11.2,1.22,5390.9,24.9746,-18.951,0.496
287203477,2925,1.08,1.171,1.38,1386.284,12.58,1.97,5451.0,29.1507,-9.1392,1.391
178215685,908,0.38,0.618,1.36,1411.432,11.1,1.26,6369.0,58.8231,-20.6242,0.44
407998915,12663,0.98,3.132,2.16,1441.111,12.9,0.87,4946.43,61.5086,7.8491,1.514
248445793,1114,0.56,9.621,1.81,1441.989,10.91,1.65,6346.0,77.6925,-3.5671,0.415
71048340,5200,0.73,11.617,2.64,1440.754,12.7,1.01,5476.0,66.3927,-17.4388,1.887
279494608,2040,0.91,1.793,1.22,1412.560,11.71,1.99,6233.0,59.0474,-17.0934,1.028
46432937,26522,0.92,1.440,1.17,1468.632,12.37,0.54,3673.0,83.869,-14.5973,1.258
317924729,10367,1.56,2.002,2.88,1468.957,11.14,1.47,7457.0,89.8631,-12.5134,
468950016,3207,1.13,8.910,4.82,1494.862,11.11,1.73,5503.57,107.6713,6.5912,1.083
99341082,5683,1.32,13.262,2.88,1497.206,11.77,1.56,5800.0,108.0372,-19.242,0.811
415090780,5127,1.61,1.747,1.80,1492.861,11.72,1.94,7107.0,114.5831,-22.6324,1.621
385921743,12760,1.80,2.336,3.12,1492.141,11.68,1.49,6982.0,111.3411,-14.7627,0.913
453054952,2469,1.31,5.238,5.04,1521.062,12.84,1.93,6219.93,145.4869,0.1214,1.99
310149582,2875,1.17,2.756,3.12,1545.578,10.88,2.03,9915.0,152.3454,-51.2615,
73644471,8180,1.30,4.591,3.84,1574.479,12.4,1.43,5596.1,191.8,-43.8404,1.322
334981118,5322,1.49,5.587,4.79,1576.932,12.24,1.87,6402.0,184.8222,-49.872,1.52
146937007,2837,0.99,3.048,1.23,1546.953,11.19,1.63,5767.0,161.2474,-46.2936,0.353
453892431,3136,1.35,5.132,2.16,1572.053,11.98,2.29,6442.0,153.5428,-69.699,1.237
467461269,2158,1.01,7.776,2.76,1574.005,9.52,2.03,6136.0,168.6396,-60.7591,0.619
176193144,5408,1.21,2.708,1.56,1630.675,11.68,1.57,6540.0,235.5493,-38.2492,1.801
397231954,2428,0.82,17.318,4.79,1633.361,10.85,1.65,6203.0,101.6654,-84.9383,0.4
122636751,2406,1.34,13.691,4.79,1635.491,10.67,1.96,5848.0,255.378,-42.4897,1.184
275614831,6212,1.82,1.140,3.48,1657.317,11.29,2.09,7437.0,282.1311,-32.7036,1.507
75076259,3025,1.03,5.968,2.88,1631.084,10.63,1.45,6600.0,266.3704,-56.883,1.761
300381700,5163,1.05,9.072,2.41,1471.205,12.43,1.41,5999.0,111.3405,-67.0675,1.417
87775884,1716,1.45,6.874,4.56,1662.637,10.64,3.08,6161.0,273.4085,-43.1875,0.677
119226450,4632,0.66,16.747,3.82,1657.066,12.24,0.89,5582.0,275.8557,-57.7873,1.258
426122272,13968,1.17,2.942,2.88,1685.132,12.37,1.54,6059.0,287.7148,34.8043,1.272
74029268,2748,1.59,7.909,5.76,1684.086,10.16,2.73,7073.0,295.2316,16.6887,1.758
230494128,6043,1.15,1.344,1.22,1684.034,9.95,1.45,5699.0,307.4905,35.8771,0.338
65612701,805,0.47,10.181,4.44,1686.254,8.92,1.74,5861.0,310.4741,36.339,0.174
364561528,14321,1.17,12.781,3.12,1713.814,12.06,0.89,6230.0,310.7289,50.3898,0.59
368100256,5673,0.64,16.151,4.44,1768.194,11.84,0.82,5783.0,356.9943,65.9618,1.103
421628129,2855,1.13,5.861,1.92,1767.682,10.76,2.02,7050.0,324.6696,58.6772,0.442
186762886,4674,1.51,7.302,3.24,1792.961,10.77,2.11,5918.82,46.3744,38.2015,0.547
28830523,6803,1.72,3.222,4.32,1792.464,10.83,2.11,7403.87,49.3697,29.5028,1.2
198490746,8673,1.18,0.904,0.96,1683.758,11.65,1.21,5212.0,196.4034,72.2557,0.524
459210578,2477,0.67,1.862,1.71,1684.746,11.92,1.35,6256.0,149.1707,81.8554,1.018
142441071,9841,0.93,2.143,2.16,1685.515,12.22,0.9,5874.0,145.7844,67.4306,1.184
86303607,7058,0.82,3.001,2.52,1843.545,12.95,0.97,4729.0,142.7067,62.2901,
21441139,2518,0.92,11.347,4.44,1878.852,11.76,1.83,5365.27,135.3358,36.2025,0.602
355792348,3081,0.81,6.388,2.88,1962.123,10.5,1.43,6634.0,230.3519,21.0107,0.486
263723967,1867,0.68,8.002,2.77,1988.135,10.11,1.55,5888.0,250.4149,15.2005,0.223
232060270,2128,0.73,17.422,4.08,1327.161,12.2,1.34,5782.0,339.1726,-61.2695,0.814
362548828,4632,1.52,7.457,5.52,1631.901,11.76,2.01,6595.0,246.124,-70.7556,1.802
308692360,4329,1.13,10.021,4.32,1525.015,12.11,1.55,5776.75,135.1819,-15.7807,1.206
73232401,1426,0.36,11.018,3.36,1593.009,10.2429,0.938119,5567.0,190.244,-43.1898,0.301
441012828,9314,1.67,2.786,3.36,1327.212,12.6438,1.63733,5740.0,323.4861,-21.1377,1.284
184728968,9829,0.90,5.947,2.77,2119.861,13.3399,0.910239,5193.0,11.9536,-11.9591,2.113
\end{filecontents*}
\csvreader[head to column names, late after line=\\] {candidates/periodictransitcandidatepart1.csv}{}{
            \TICID & \epoch & \radius & \tdepth & \orbitalperiod & \duration
        }
    \end{tabular}
\end{table}

\begin{table}
    \centering 
    \caption*{(Continued) List of single-planet multi-transit light curve candidates.}
    \begin{tabular}{lccccc} 
        \hline 
        \hline
        TIC ID & Transit & Radius & Depth & Orbital& Duration \\ 
        & Epoch & (Jupiter &  & Period& \\
        &  & radii) & (ppm) & (days)& (hours)\\
        \hline
        \begin{filecontents*}{candidates/periodictransitcandidatepart2.csv}
TICID,tdepth,radius,orbitalperiod,duration,epoch,TessMag,radiustarget,temperaturetarget,RA,DEC,RVerror
328100489,10920,1.53,4.702,4.08,1389.925,11.9524,1.47104,5683.0,14.3051,-12.0135,2.485
305540712,5084,0.78,15.082,2.40,1522.485,12.4298,1.07428,6127.0,139.483,-16.3105,1.954
365717150,754,0.99,11.907,3.48,1608.829,9.5244,2.32258,6422.0,210.8984,-32.5542,0.428
383829317,2421,1.06,9.912,2.64,1689.937,11.2722,2.10036,6576.0,284.1435,33.503,1.078
355206662,7142,0.77,3.934,1.20,1874.296,12.3906,0.90576,5156.0,133.4095,46.6562,1.065
137020190,6281,0.84,23.057,3.36,1901.809,12.4379,1.06118,5873.0,197.0925,44.2026,1.238
459801869,3004,1.05,20.381,4.81,1933.236,9.19,1.91231,6875.03,196.69,38.989,0.375
436255371,2588,0.88,24.028,3.84,1703.092,11.5906,1.56119,7128.0,306.6309,24.7434,1.298
321011198,4333,0.48,7.402,2.52,1388.179,13.173,0.757324,4811.0,43.5787,-41.1811,2.228
168790078,1511,0.63,6.342,1.44,1416.657,11.0476,1.70985,6589.0,63.4807,-32.9861,0.757
250133183,4155,1.02,4.478,1.44,1442.901,12.3841,1.57645,5884.41,63.2176,-3.5007,1.302
386518121,6306,0.72,6.051,1.92,1524.317,12.5665,1.05338,5409.0,144.272,-14.2545,1.509
185980211,1191,0.71,2.962,2.28,1521.703,10.4556,1.86939,5601.0,131.1926,-34.4444,0.862
147566325,1596,1.41,8.285,3.62,1548.537,10.997,2.62066,6089.0,163.596,-42.3912,0.292
4809705,1908,0.57,4.461,3.12,1549.352,10.9528,1.08452,5845.22,141.4404,-36.0163,0.541
163029327,5454,0.72,2.562,1.61,1574.351,12.3648,0.949331,5415.0,191.0291,-48.1029,1.272
20614436,2625,0.52,7.601,1.25,1574.608,11.1105,1.09541,5743.0,201.3705,-19.001,0.34
275548876,881,0.55,9.053,2.16,1634.489,9.2829,1.75692,6257.0,240.5557,-53.1416,0.197
35771052,1819,1.06,7.059,2.76,1630.969,10.4858,2.19503,5829.0,254.1372,-34.4674,0.769
85651189,1898,0.81,4.081,2.88,1663.123,10.7222,1.63732,5612.0,271.2142,-42.3673,0.389
251619073,1531,0.88,1.578,2.52,1791.553,9.26414,3.19833,6271.0,45.859,55.389,0.88
158922455,3671,0.45,1.633,1.86,1685.843,12.1479,0.72009,4563.5,288.823,42.5427,1.007
295117707,5393,0.76,6.881,1.92,1719.467,12.0264,0.958166,5764.0,303.0297,50.428,1.114
193386434,2628,1.28,2.868,2.76,1687.317,11.8827,2.07993,6665.0,305.6938,44.44,1.212
63637169,891,0.99,5.622,4.08,1741.783,10.8531,3.21927,7445.0,320.1529,53.3748,1.022
388076435,619,0.29,2.565,1.30,1742.662,9.17559,1.13741,6043.0,317.8838,68.4115,0.214
73638456,2475,0.48,2.108,2.16,1793.813,11.294,0.955791,5631.0,39.6649,35.0136,0.881
428320381,6016,1.80,2.434,4.08,1794.662,11.1321,2.52723,7410.0,55.2602,52.1362,0.568
273858602,3855,0.64,1.619,1.44,1819.868,12.1702,0.98559,5595.0,86.5414,43.5373,0.979
350215049,4360,0.83,9.124,4.44,1333.385,12.8183,1.25273,6162.93,355.7447,-60.9082,2.225
152508218,3918,0.66,9.305,2.52,1415.589,12.8886,1.03444,5815.0,63.5112,-42.3633,2.67
117646374,2641,0.62,1.889,1.84,1439.241,12.8247,1.14934,5376.0,67.899,-14.1289,2.135
138854818,9692,1.78,5.789,4.82,1472.193,12.0727,1.82625,6509.96,82.8035,5.0349,2.648
150165026,5709,1.25,7.778,1.56,1414.126,12.553,1.66176,6615.0,93.1517,-61.7784,2.144
32164949,20882,0.57,8.463,2.88,1496.752,13.2878,0.399893,3559.0,117.0253,-8.6674,2.559
463279737,1589,1.12,4.961,2.16,1547.778,10.6546,2.59847,6209.0,153.4949,-55.7424,2.399
410056850,6915,1.13,10.618,4.44,1664.842,12.6847,1.34041,6543.0,320.8737,-71.2386,2.928
229539379,8450,1.68,9.644,4.81,1685.014,12.8307,1.88021,5594.0,222.1919,74.0447,2.174
391839433,7637,1.69,2.721,3.72,1816.503,11.2528,1.86053,6947.83,72.1517,42.5511,2.14
193034981,7332,1.67,3.547,3.48,1686.796,12.0125,1.76495,6062.0,304.9101,42.6974,1.625
425937489,7491,1.13,15.572,1.19,1326.472,12.5121,1.26525,5591.0,5.3352,-62.9972,5.99
38833595,4475,1.13,12.043,1.22,1327.242,11.857,2.01658,6065.0,7.8422,-68.5209,6.218
314867507,6987,1.18,2.474,1.89,1354.861,12.8219,1.40874,5636.22,50.7275,-78.4566,3.823
343813363,10611,1.10,5.011,3.48,1414.026,13.1723,1.07022,5841.0,49.9088,4.3632,3.033
439902212,10527,1.18,22.393,4.19,1412.141,12.5967,1.14434,6129.0,44.924,-0.7581,3.613
407993292,11509,1.28,5.977,1.23,1442.994,12.7642,1.19318,5785.79,60.9587,11.1784,3.449
187102522,6852,1.78,2.149,3.48,1519.277,11.1692,2.1688,9432.0,134.6482,-34.748,5.152
146320244,1910,0.54,4.485,2.52,1544.876,10.8867,1.07596,5394.0,158.4684,-42.6995,0.312
137391077,3450,1.15,9.032,4.44,1989.352,12.0836,2.05734,7206.0,52.1749,73.8069,4.343
341938350,3117,0.87,14.027,3.48,2010.969,11.6016,1.33216,6277.0,268.2166,23.2657,0.785
258243070,9345,1.77,7.462,4.08,1985.017,12.6523,1.77717,5496.61,252.7222,26.396,2.435
339502646,2897,0.88,15.625,3.72,1847.148,11.9704,1.59136,6097.0,92.538,64.8818,0.861
39015621,4350,0.62,6.102,3.72,1845.614,12.3181,0.903417,5487.66,118.4733,38.0627,1.908
87878954,2070,0.94,2.314,3.12,1817.800,11.8049,1.92957,6383.45,81.9719,45.6506,
242843507,14525,1.74,19.303,4.08,1765.785,12.6403,1.43617,6345.0,9.2105,20.836,3.192
373662662,1736,0.78,6.924,2.76,1740.749,11.1204,1.70726,7632.0,320.5451,39.515,1.112
64770659,1942,0.48,4.607,3.72,1713.666,11.4231,0.847053,4879.0,322.5151,53.8565,1.283
230966894,4245,1.13,2.651,1.44,1655.421,12.518,1.67943,6309.0,287.8733,-54.997,2.85
253637358,4976,1.33,3.451,2.52,1660.381,12.3124,1.68661,5800.0,281.34,-38.0479,3.531
193164274,11024,1.79,4.012,3.12,1629.370,11.1016,1.21351,5215.0,259.73,-38.0304,
\end{filecontents*}
\csvreader[head to column names, late after line=\\] {candidates/periodictransitcandidatepart2.csv}{}{
            \TICID & \epoch & \radius & \tdepth & \orbitalperiod & \duration
        }
        \hline 
    \end{tabular}
\end{table}

\begin{table}
    \centering 
    \caption{List of single-transit light curve candidates.}
    \label{tab:singletransit_candidates}
    \begin{tabular}{lccccc} 
        \hline 
        \hline
        TIC ID & Transit & Radius & Depth & RV Error & Sector \\ 
        & Epoch & (Jupiter &  & & \\
        &  & radii) & (ppm) & $\text{km s}^{-1}$ & \\
        \hline 
        \begin{filecontents*}{candidates/singletransitercandidatepart1d.csv}
TICID,epoch,radius,tdepth,RVerror,TessMag,radiustarget,temperaturetarget,RA,DEC,sector,duration
52285162,1352.935,0.70,6540,1.231,12.39,0.86,5490.0,21.6936,-66.0828,1,4.68
89546401,1363.241,1.27,7385,1.354,12.51,1.48,5619.0,358.1803,-23.1570,2,8.85
358289302,1354.862,0.70,5365,1.664,12.44,0.95,5874.0,62.1765,-57.5948,2,4.03
199735531,1357.684,0.92,5317,1.287,12.11,1.55,5699.0,22.5193,-46.7577,2,1.82
122522333,1389.3684,0.74,1792,0.41,10.71,1.76,6230.0,38.5830,-31.8483,3,6.78
63921468,1400.3114,0.86,1256,0.419,8.88,2.44,6429.0,13.2707,-24.9517,3,9.54
100991578,1433.509,0.65,1313,0.544,10.83,1.8,6290.0,54.8863,-45.8695,4,13.97
343839315,1415.543,1.15,5771,1.692,12.35,1.43,5850.47,52.0129,3.7024,4,3.29
9939655,1416.462,0.87,1766,0.283,9.21,2.56,6432.35,49.7325,-7.0181,4,3.31
92304420,1413.747,0.44,2832,0.322,10.85,0.82,4927.0,50.7078,-18.7711,4,3.67
283631762,1457.599,0.69,4708,0.642,11.59,1.01,5893.25,62.1539,8.3620,5,4.82
123570723,1468.862,0.99,5688,0.41,11.32,1.26,5798.0,88.9923,-21.2054,6,7.03
438122862,1475.322,0.93,2071,2.467,10.3204,2.03,7298.0,95.9343,12.9431,6,16.15
231340540,1495.131,0.74,5172,0.908,12.11,0.99,5709.84,97.0225,-30.9286,7,4.66
149965576,1499.412,1.23,3243,0.309,10.61,2.07,5713.0,116.7289,-30.2755,7,6.72
388014300,1499.69,1.45,10440,1.6,12.39,1.41,6273.63,120.3558,10.3218,7,11.64
455201271,1494.362,0.71,5309,0.742,11.59,0.94,5655.0,125.0583,1.6217,7,7.01
141786309,1507.399,1.44,2300,1.006,11.79,2.65,6270.0,117.7526,-18.1192,7,2.45
434452008,1539.751,0.62,2754,0.456,11.41,1.1,5743.0,139.7381,-19.2334,8,7.44
117888314,1526.051,0.78,1083,1.152,11.57,2.31,6756.0,128.2155,-20.6270,8,4.32
19265777,1520.933,0.81,2083,0.765,10.06,1.72,7924.0,139.0626,-33.5293,8,10.46
238060327,1519.101,1.30,7427,0.95,11.81,1.5,5665.0,104.1960,-52.0740,8,8.59
334656125,1523.374,0.70,1782,0.353,10.26,1.65,6705.0,141.3828,-14.4917,8,4.44
289596640,1539.431,0.91,2168,0.247,10.68,2.33,5034.0,131.7130,-13.4607,8,4.27
106056929,1548.591,0.83,5170,1.138,11.82,1.13,5963.16,157.7914,-37.6156,9,10.39
36541001,1566.125,1.25,4253,0.779,11.67,1.81,6426.0,150.4157,-43.8487,9,9.07
147027702,1550.517,1.21,3022,0.331,10.63,2.1,6946.0,161.7729,-43.9318,9,11.59
105712339,1561.807,0.89,6512,1.212,12.2,1.12,5858.16,156.4123,-38.8594,9,2.04
40561422,1593.072,1.12,2848,0.904,11.48,2.1,5773.0,185.9559,-12.1095,10,5.76
97975310,1577.598,1.09,4686,0.433,11.1,1.59,6316.0,199.8431,-29.7846,10,8.26
57533297,1590.158,0.93,7343,1.382,12.56,1.07,5793.0,173.2094,-33.3155,10,5.69
266376971,1586.723,0.30,778,0.183,8.28,1.07,5966.93,175.1695,-58.1043,10,2.21
394354824,1620.739,1.67,6138,1.616,12.37,1.96,5686.0,152.0954,-81.8301,11,5.47
449486364,1642.242,1.35,5876,0.966,11.35,1.65,6490.79,160.6103,-72.6441,12,12.17
247934029,1646.182,0.42,1169,-,8.9,1.16,5778.0,254.4464,-43.7998,12,6.89
182456927,1670.689,1.03,5352,0.879,11.63,1.17,6023.0,281.8916,-46.3894,13,11.52
265371495,1659.879,0.92,7616,1.418,12.46,1.05,5643.0,315.5682,-73.7537,13,6.17
29778074,1660.486,0.75,3384,1.423,11.85,1.28,6133.55,64.4541,-66.3704,13,8.16
42196078,1697.922,0.85,3002,0.287,10.52,1.51,5883.0,286.3133,31.8138,14,8.86
210175389,1690.582,1.77,15190,1.47,12.01,1.41,6514.0,307.1381,20.5063,14,6.22
48304182,1688.822,1.35,6803,1.114,12.04,1.4,6392.16,283.1546,48.9070,14,5.26
260908297,1686.922,0.71,5533,0.466,11.44,0.9,5659.0,303.5420,23.2550,14,3.58
147898114,1698.252,1.06,9752,1.477,12.71,1.05,5491.0,164.7757,67.1008,14,2.52
233577004,1704.192,1.16,5721,0.354,11.47,1.47,5810.0,282.3459,64.2780,14,8.83
269130683,1698.191,1.06,4149,0.931,10.78,1.55,6657.0,299.6243,42.8044,14,4.78
341687821,1692.988,0.56,1752,0.315,10.24,1.35,6204.0,192.2733,66.8937,14,9.62
194125303,1705.256,0.27,988,0.204,9.67,0.83,5529.0,307.2263,44.5445,14,3.43
224299778,1727.366,0.88,1947,0.3,10.59,2.0,5507.52,185.6179,57.1782,15,7.99
282926264,1731.746,1.52,7931,1.27,12.56,1.71,5630.0,226.8549,60.5525,15,4.15
219466517,1731.474,0.97,1937,0.226,9.68,2.16,6054.0,209.5843,75.5635,15,11.95
256783400,1733.189,0.97,3983,1.251,10.95,1.43,6294.0,329.3189,36.0532,15,3.89
429219512,1730.989,1.44,5455,1.336,11.63,1.85,6499.0,313.2470,57.4798,15,12.14
288160753,1730.683,0.91,11791,1.939,12.63,0.81,5067.0,221.2818,77.9309,15,3.31
99919551,1719.081,0.68,8121,1.089,12.62,0.75,4387.0,169.1384,60.0147,15,3.05
174307526,1757.823,0.60,18868,1.822,13.11,0.44,3611.0,352.5884,46.2899,16,3.84
162033431,1743.642,0.61,3361,0.575,11.45,1.03,5613.0,330.6288,30.2897,16,3.86
412119944,1779.572,0.99,3902,0.783,11.6,1.51,6254.0,329.1712,62.8392,17,11.23
85254937,1785.362,0.99,3813,0.664,11.03,1.6,6112.25,21.6725,33.9389,17,7.63
440675630,1766.572,0.85,5585,0.724,11.88,1.23,5770.29,8.2251,24.8235,17,10.73
241134937,1796.731,1.01,4899,0.25,10.72,1.42,5969.99,30.2782,48.2855,18,10.99
\end{filecontents*}
\csvreader[head to column names, late after line=\\] {candidates/singletransitercandidatepart1d.csv}{}{
            \TICID & \epoch & \radius & \tdepth & \RVerror & \sector
        }
    \end{tabular}
\end{table}

\begin{table}
    \centering 
    \caption*{(Continued) List of single-transit lighr curve candidates.}
    \begin{tabular}{lccccc} 
        \hline 
        \hline
        TIC ID & Transit & Radius & Depth & RV Error & Sector \\ 
        & Epoch & (Jupiter &  & & \\
        &  & radii) & (ppm) & $\text{km s}^{-1}$ & \\
        \hline 
        \begin{filecontents*}{candidates/singletransitercandidatepart2d.csv}
TICID,epoch,radius,tdepth,RVerror,TessMag,radiustarget,temperaturetarget,RA,DEC,sector,duration
397986100,1799.102,1.10,7548,0.765,11.66,1.35,6314.0,23.4877,71.8234,18,8.57
326808653,1810.721,1.12,2085,-,10.26,2.15,6272.0,22.4760,47.4914,18,7.3
402848417,1829.641,1.21,5988,1.452,12.2,1.54,5888.0,17.7422,82.2215,19,5.14
82011491,1830.911,0.48,773,0.179,9.21,1.72,6378.0,48.8777,60.6347,19,6.02
233096001,1823.721,0.72,5507,1.261,12.24,0.97,5398.0,273.0947,65.2112,19,3.29
273856464,1822.941,0.82,3080,0.86,10.93,1.42,6605.76,86.6032,42.2518,19,8.16
252895642,1853.361,0.68,1927,0.251,9.87,1.55,6373.0,103.5033,49.8633,20,14.16
232616346,1853.731,1.26,4006,0.428,11.53,1.9,5438.0,259.3617,63.4543,20,12.0
68380016,1848.051,1.13,2019,-,10.72,2.5,6490.0,102.7268,35.2242,20,16.32
229702978,1874.501,1.21,5275,1.429,11.73,1.67,6151.0,277.5597,67.6013,21,11.97
302728777,1877.541,0.91,4267,0.842,12.32,1.26,5569.11,169.5089,54.2824,21,5.86
85334736,1906.357,0.71,3180,0.313,10.67,1.26,5778.05,168.1713,31.8053,22,9.02
144314863,1906.027,0.62,2460,0.446,10.86,1.25,6009.91,170.6618,35.3080,22,6.38
115861501,1919.057,0.53,3378,0.812,11.54,0.9,5207.69,175.0901,43.4430,22,3.74
207363637,1922.867,0.74,10753,1.38,12.6,0.71,4466.0,208.5341,47.6075,22,3.26
458457997,1938.378,1.15,12857,1.998,12.65,1.01,5945.0,209.6756,13.2078,23,4.2
416023266,1937.798,1.06,4697,1.036,12.1,1.49,5781.0,209.1109,18.0507,23,11.4
135221230,1936.528,0.82,6452,0.505,11.95,1.01,5448.21,214.6029,22.3766,23,6.12
41743658,1960.408,0.98,2469,0.148,9.32,1.97,5489.0,233.1976,37.9219,24,9.0
18632388,2005.602,1.21,3035,-,10.46,2.0,5507.0,266.8006,34.3859,25,10.34
357597620,1993.972,1.25,7207,1.462,12.12,1.47,6476.0,264.4840,28.7935,25,5.47
275267824,2020.552,1.36,7222,0.405,11.15,1.43,6283.07,269.0391,22.7469,26,4.99
\end{filecontents*}
\csvreader[head to column names, late after line=\\] {candidates/singletransitercandidatepart2d.csv}{}{
            \TICID & \epoch & \radius & \tdepth & \RVerror & \sector
        }
        \hline 
    \end{tabular}
\end{table}

\begin{table}
    \centering
	\vspace{5mm}
	\caption{List of candidates identified as multi-planet system.}
	\label{tab:multiplanet_candidates}
	\vspace{4mm}
	\begin{tabular}{lcccc} 
		\hline
		\hline
        TIC ID  & TessMag  & Transit & Depth &  Radius\\
		 & & Epoch& (ppm) & (Jupiter\\
		&&&& radii)\\
        \hline
        TIC 193096383 (b)&12.23&1546.82&2870&0.57\\
        TIC 193096383 (c)&12.23&1559.01&9564&1.04\\
        TIC 221567884 (b)&9.09&1629.92&868&0.48\\
        TIC 221567884 (c)&9.09&1648.66&2169&0.75\\
        TIC 221567884 (d)&9.09&2372.65&3037&0.91\\
        TIC 118798035 (b)&11.29&1656.35&5970&0.86\\
        TIC 118798035 (c)&11.29&1663.91&10660&1.16\\
        TIC 400049903 (b)&11.61&1669.72&4041&0.61\\
        TIC 400049903 (c)&11.61&1778.60&2424&0.47\\
        \hline
        \hline		
	\end{tabular}
	\vspace{3mm}
\end{table}

\begin{figure}
\begin{center}
\vspace{5mm}
\includegraphics[width=0.48\textwidth]{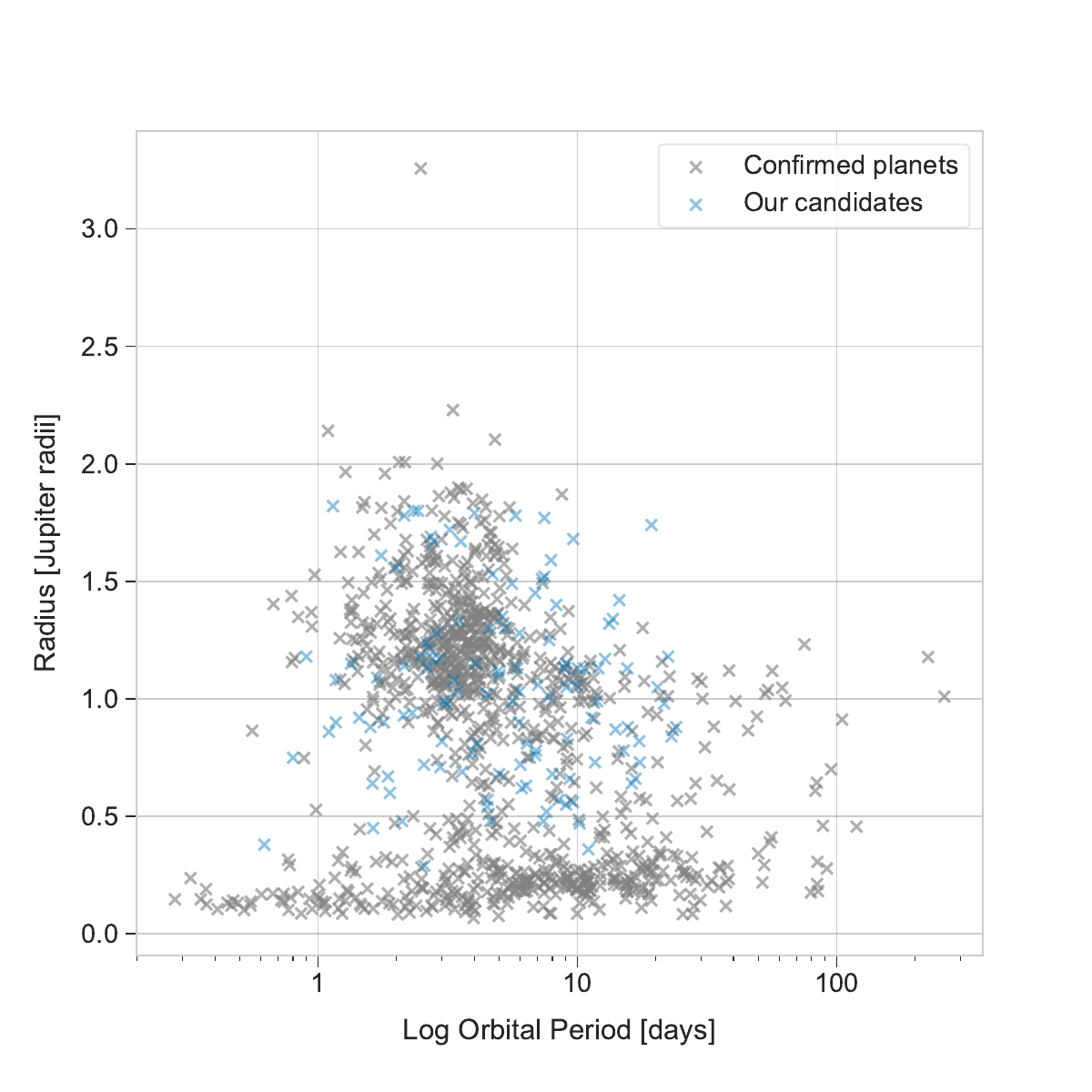}
\vspace{-2mm} 
\caption{\label{fig:log_radius_vs_orbital_period}
Relationship between radius and log orbital period. Each point represents a confirmed planet from ExoFOP-TESS or our candidates.}
\vspace{-1mm}
\end{center}
\end{figure}

%% file: conclusions.tex
\section{Conclusions}
\label{sec:conclusions}

This work demonstrates the Transformer's ability to capture long-range dependencies and focus on relevant segments of light curves, combined with a CNN embedding that effectively aids the encoder in learning short-range patterns and variations. This combination allows for the detection of both multi-transit and single-transit light curves without prior transit parameters. Moreover, incorporating background and centroid time series data significantly enhanced the performance of our model, allowing it to better represent variations in the light curves, including stellar variability and instrumental effects. Although the use of Transformers in exoplanet detection, and in astronomy in general, is still in its early stages, the success of this approach suggests that Transformers could offer a valuable alternative to traditional methods.  \par


Based on our findings, our proposed model learned the signal of an exoplanet transit independently of its periodicity, allowing it to identify transit events even in the absence of regular repeating patterns. This ability to generalize beyond periodic signals is particularly important for detecting single transiters, where only one transit event is observed due to long orbital periods. However, through our analysis, we found that our model did not identify any Earth-sized or super-Earth candidates. Instead, it successfully detected exoplanets with radii greater than 0.25 Jupiter radii. The Earth-sized and super-Earth planets typically produce shallower transits, which makes their detection more reliant on periodicity. In these cases, the recurring transit events allow traditional period-searching algorithms, such as BLS, to identify the smaller, more subtle signals that our model may not yet effectively detect. In addition to detecting multi-transit light curve candidates with radii greater than 0.25 Jupiter radii, our model also effectively identified single transiters and multi-planet systems. In total, we identified 214 planetary system candidates, which include 122 single-planet multi-transit light curve candidates, 88 single-transit candidates and 4 associated with multi-planet systems. \par


As a result, the strength of our NN is that it is able to detect exoplanet transit signals within light curves without relying on their periodicity. This is achieved through training the NN to recognize the distinctive shape of the planetary transit signals, even in the presence of stellar variability. While the detection of smaller, less prominent planets remains a challenge, the success in identifying long-period candidates and complex systems underscores the potential of Transformer-based models in expanding the scope of exoplanet detection. In particular, our model operates without the need for prior transit parameters or phase folding, which are typically required by the previous approach of DL, enabling a more direct analysis of the data. This capability allows for the identification of diverse transit signals that may be overlooked by conventional techniques.  \par

Our future work aims to analyze data from TESS sectors $>$ 26 to identify new candidates. Additionally, we plan to improve our detection models to achieve more accurate identification of smaller exoplanets, such as Earth-sized and super-Earth candidates, while maintaining our capability to detect transits regardless of their periodicity to detect multi-planet systems or single transiters. We also consider that upcoming missions, such as PLATO \citep{rauer2016plato} and the Nancy Grace Roman Space Telescope Roman, will present new challenges in the exoplanet search. The Roman mission, in particular, is predicted to discover between $\sim$60,000 and $\sim$200,000 transiting planets \citep{wilson2023transiting}. The approach we proposed provides a base for the development of new architectures.

%% file: appendix.tex
In this section, we describe the Multi-Head Self-Attention (MSA) mechanism used in the encoder. 

Given an input sequence of embeddings $\mathbf{Z}_{l-1}$ where $T$ is the sequence length and $d$ is the dimensionality of the embeddings, the self-attention mechanism computes a set of attention scores. It is computed using three vectors, the queries $\mathbf{Q}$, keys $\mathbf{K}$, and values $\mathbf{V}$. Vectors $Q$ and $K$ are compared to calculate similarities and obtain scores/weights for vector $V$. To compute these vectors as three different subspace representations, the input embeddings is projected with learned weight matrices defined as:

\begin{equation}
\begin{split}
\mathbf{Q=Z}_{l-1}\mathbf{W^{\mathnormal{Q}} \in \mathbb{R}^{\mathnormal{n \times d_q}}},\\
\mathbf{K=Z}_{l-1}\mathbf{W^{\mathnormal{K}} \in \mathbb{R}^{\mathnormal{n \times d_k}}},\\
\mathbf{V=Z}_{l-1}\mathbf{W^{\mathnormal{V}} \in \mathbb{R}^{\mathnormal{n \times d_v}}}.
\label{eq:attention_qkv}
\end{split}
\end{equation}

where $\mathbf{W}^{\mathnormal{Q}}$, $\mathbf{W}^{\mathnormal{K}}$, $\mathbf{W}^{\mathnormal{V}}$ are the learned weight matrices, and their dimensionalities are $d \times d_k$, with $d_k$ represents the dimensions of the queries and keys. 

Next, we calculate the attention scores using the scaled dot-product, self-attention matrix $\mathbf{A} \in \mathbb{R}^{n \times d_v}$ can be expressed as:

\begin{equation}
\mathbf{A}=\mathrm{Attention\mathbf{(Q,K,V)} = softmax \left ( \frac{\mathbf{QK}^\top}{\sqrt{d_k}} \right ) \mathbf{V}}
\label{eq:scaled-dot-product}
\end{equation}

The attention layer takes the input in the form of those three parameters: queries $\mathbf{Q}$, keys $\mathbf{K}$ and values $\mathbf{V}$. These three parameters have similar structure, where each element in the sequence is represented by a vector. The query, key, and value vectors are derived from the input embeddings and they are linearly projected $h$ times, allowing their computations to be repeated multiple times in parallel \citep{vaswani2017attention}. The outputs of these individual heads are concatenated and passed through a linear projection to create the final output defined by:

\begin{equation}
\label{multihead_eq}
\mathrm{MSA}(\mathbf{Z}_{l-1}) = \mathrm{MultiHead\mathbf{(Q,K,V)} = 
\mathbf{W}^O \begin{bmatrix}
 h_1\\
.\\
.\\
.\\
 h_h
\end{bmatrix}
}
\end{equation}

with learned output weights $\mathbf{W}^O \in \mathbb{R} ^{hd_v\times d} $, and each attention head $h_i$ computes the attention as follows:

\begin{equation}
    h_i = \mathrm{Attention({\mathbf{Q}}_i, {\mathbf{K}}_i,{\mathbf{V}}_i) }
\end{equation}

The resulting multi-head attention output is then passed through a residual connection and layer normalization:

\begin{equation}
\mathbf{Z'}_{l} = \mathrm{MSA}({\mathrm{Norm}(\mathbf{Z}_{l-1})}) + {\mathbf{Z}_{l-1}}, \quad \mathrm{for} \quad l=1,..., L
\end{equation}
